\documentclass[a4paper,fleqn,usenatbib]{mnras}

\usepackage{newtxtext,newtxmath}
\usepackage{soul}

\usepackage[T1]{fontenc}
\usepackage{ae,aecompl}

\usepackage{graphicx}	
\usepackage{amsmath}	
\usepackage{amssymb}	
\usepackage{caption}
\usepackage{subfigure}
\usepackage{xcolor}
\usepackage{sansmath}

\definecolor{dg}{rgb}{0,0,0}

\title[Turbulence profiling using AO telemetry]{Optimising the accuracy and efficiency of optical turbulence profiling using adaptive optics telemetry for extremely large telescopes}

\author[D. J. Laidlaw et al.]{Douglas J. Laidlaw,$^{1}$\thanks{E-mail: douglas.j.laidlaw@durham.ac.uk}
James Osborn,$^{1}$
Timothy J. Morris,$^{1}$
Alastair G. Basden,$^{1}$ \newauthor
Olivier Beltramo-Martin,$^{2}$
Timothy Butterley,$^{1}$
Eric Gendron,$^{3}$
Andrew P. Reeves,$^{1,4}$\newauthor
G\'erard Rousset,$^{3}$
Matthew J. Townson,$^{1}$
and Richard W. Wilson$^{1}$\\
$^{1}$Centre for Advanced Instrumentation (CfAI), Durham University, South Road, Durham, DH1 3LE, United Kingdom. \\
$^{2}$Aix Marseille Univ., CNRS, CNES, LAM, 38 rue F. Joliot-Curie, 13388 Marseille Cedex 13, France.\\
$^{3}$LESIA, Observatoire de Paris, Universit$\acute{e}$ PSL, CNRS, Sorbonne Universit$\acute{e}$, Univ. Paris Diderot, \\Sorbonne Paris Cit$\acute{e}$, 5 place Jules Janssen, 92195 Meudon, France.\\
$^{4}$Institute of Communication and Navigation, German Aerospace Center (DLR), 82234, Germany.}

\date{Accepted XXX. Received YYY; in original form ZZZ}

\pubyear{2016}

\begin{document}
\label{firstpage}
\pagerange{\pageref{firstpage}--\pageref{lastpage}}
\maketitle

\begin{abstract}
Advanced adaptive optics (AO) instruments on ground-based telescopes require accurate knowledge of the atmospheric 
turbulence strength as a function of altitude. This information assists point spread function reconstruction, AO 
temporal control techniques and is required by wide-field AO systems to optimise the reconstruction of an observed 
wavefront. The variability of the atmosphere makes it important to have a measure of the optical turbulence profile in 
real-time. This measurement can be performed by fitting an analytically generated covariance matrix to the 
cross-covariance of Shack-Hartmann wavefront sensor (SHWFS) centroids. In this study we explore the 
benefits of reducing cross-covariance data points to a covariance map region of interest (ROI). 
{\color{dg}A technique for using the covariance map ROI to measure and compensate for SHWFS misalignments is also introduced. 
We compare the accuracy of covariance matrix and map ROI optical turbulence profiling using both simulated and 
on-sky data from CANARY, an AO demonstrator on the 4.2\,m William Herschel telescope, La Palma. 
On-sky CANARY results are compared to contemporaneous profiles from 
Stereo-SCIDAR - a dedicated high-resolution optical turbulence profiler. It is shown that the covariance map ROI 
optimises the accuracy of AO telemetry optical turbulence profiling. In addition, we show that the covariance map ROI reduces 
the fitting time for an extremely large telescope-scale system by a factor of 72. 
The software package we developed to collect all of the 
presented results is now open-source.}

\end{abstract}




\section{Introduction}
\label{sec:int}

\begin{table*}
    \centering
    \caption{Physical parameters of CANARY, AOF and ELT-scale AO systems.}
    \label{tab:params}
    \begin{tabular}{|l|c|c|c|c|}
    \hline
    AO System & Pupil Diameter (m) & SHWFS Dimensions (Sub-apertures)  & Sub-apertures per SHWFS & Sub-aperture Baselines per SHWFS \\ \hline
    CANARY    & $4.2$  & $7\times7$       & $36$      & $129$ \\ \hline
    AOF       & $8.2$  & $40\times40$      & $1240$     & $4989$         \\ \hline
    ELT-scale   & $39.0$  & $74\times74$      & $4260$    & $17101$     \\ \hline
    \end{tabular}
\end{table*}

Advanced ground-based optical telescopes can employ adaptive optics (AO) systems to compensate for 
wavefront perturbations. The primary cause of these aberrations are 
atmospheric refractive index fluctuations. AO systems are able to detect the strength of phase 
perturbations across a field of view (FoV) by utilising Shack-Hartmann wavefront sensors (SHWFSs). This information is then relayed to 
one or more deformable mirrors (DMs) - situated in the optical path - that act to mitigate the measured perturbations.
To achieve high angular resolution in wide-field astronomy AO systems must be capable of tomographically 
reconstructing the wavefront across a large FoV. This technique often requires an estimation of the optical turbulence profile i.e. a measure 
of the refractive index structure constant, $C_{n}^{2}(h)$, where $h$ is altitude. 
The optical turbulence profile directly impacts wavefront correction in wide-field AO (\citealp{Neichel2009}). 
Precise point spread function (PSF) reconstruction is therefore
reliant on the accuracy of its measurement (\citealp{Villecroze2012}; \citealp{Martin2016}). 
Additional applications for optical turbulence profiling 
include AO temporal control techniques (\citealp{Petit2009}), queue scheduling of science cases and forecasting 
of atmospheric parameters (\citealp{Masciadri2012}). Forecasts may in turn be validated and 
calibrated by a subsequent measure of the optical turbulence profile.

Slope Detection and Ranging (SLODAR; \citealp{Wilson2002}) is a widely-used
technique for measuring the optical turbulence profile. It uses a number of SHWFSs 
to geometrically triangulate turbulent zones in the atmosphere. SHWFSs are used to measure wavefront 
perturbations and so the SLODAR 
technique can utilise AO telemetry. One possible SLODAR-based technique for measuring the 
optical turbulence profile is to first calculate the cross-covariance 
between the centroid measurements from all SHWFSs. This expresses the optical turbulence profile 
as a covariance matrix. The optical turbulence profile can be recovered by iteratively fitting an analytical model 
to the measured covariance matrix (\citealp{Vidal2010}). Covariance matrix 
fitting can also be used for parameter estimation e.g. measuring SHWFS misalignments (\citealp{Martin2016}). 
To assure science goals are met forthcoming extremely large telescope (ELT) AO systems must be updated regularly with an accurate 
measure of the optical turbulence profile (\citealp{Peng2018}). However, the covariance matrix of an ELT AO system will contain millions of 
cross-covariance measurements. This makes it a challenge to perform efficient ELT 
covariance matrix fitting.

{\color{dg}
By averaging covariance matrix baselines the optical turbulence profile can be studied as 
a covariance map. Previous studies have shown that a covariance map region of interest (ROI) can be used to 
perform SLODAR data analysis (\citealp{Butterley2006}; \citealp{Cortes2012}; \citealp{Guesalaga2014}; {\color{dg}\citealp{Guesalaga2017}}; \citealp{Ono2016}). 
However, an investigation into the optimal size of the covariance 
map ROI has not been published. We describe our methodology to define the covariance map ROI and we 
explore its benefits for SLODAR data analysis. It is compared to 
its covariance matrix counterpart throughout. Three different scales of AO systems are considered: 
CANARY, a multi-object AO (MOAO) pathfinder on the $4.2$\,m William Herschel
telescope (WHT), La Palma; the adaptive optics facility (AOF) on an $8.2$\,m unit telescope of 
the very large telescope (VLT), Paranal; and an instrument 
designed for the $39$\,m European ELT, Cerro Armazones. The 
parameters of each AO system are listed in Table~\ref{tab:params}.

We assess the sensitivity of the two SLODAR methods with respect to SHWFS misalignments in rotation 
and lateral shift. It was shown by \cite{Martin2016} that 
the covariance matrix can be used to measure SHWFS misalignments. If these misalignments can be measured they can 
be compensated for during the optical turbulence profiling procedure. We build upon this work and show that with a 
modified fitting procedure, the covariance map ROI can also measure SHWFS misalignments.

A qualitative study of AO telemetry optical turbulence profiling compares the accuracy of the two SLODAR methods. 
The comparison is made using both simulated and on-sky CANARY data (\citealp{Morris2014}). In additon, we 
investigate the optimal size of the covariance map ROI. Contemporaneous measurements 
made using the scintillation detection and ranging (SCIDAR; \citealp{Vernin1973}) technique were provided as a 
reference to the expected on-sky optical turbulence profile. These 
measurements came from the Stereo-SCIDAR instrument (\citealp{Shepherd2013}) on the 2.5\,m Isaac Newton 
telescope (INT), La Palma (\citealp{Osborn2015}). The Stereo-SCIDAR instrument does not utilise 
AO telemetry and instead performs optical turbulence profiling through the 
cross-covariance of atmospheric scintillation patterns. It is a dedicated stereoscopic 
high-resolution optical turbulence profiler.

Quantitative analysis is presented for the reduction in computational time when performing SLODAR data analysis with a covariance map ROI instead of a 
covariance matrix. The potential benefits for an ELT-scale instrument are considered.
}

\section{Measuring the optical turbulence profile}
\label{sec:slopeCov}

\begin{figure}
	\includegraphics{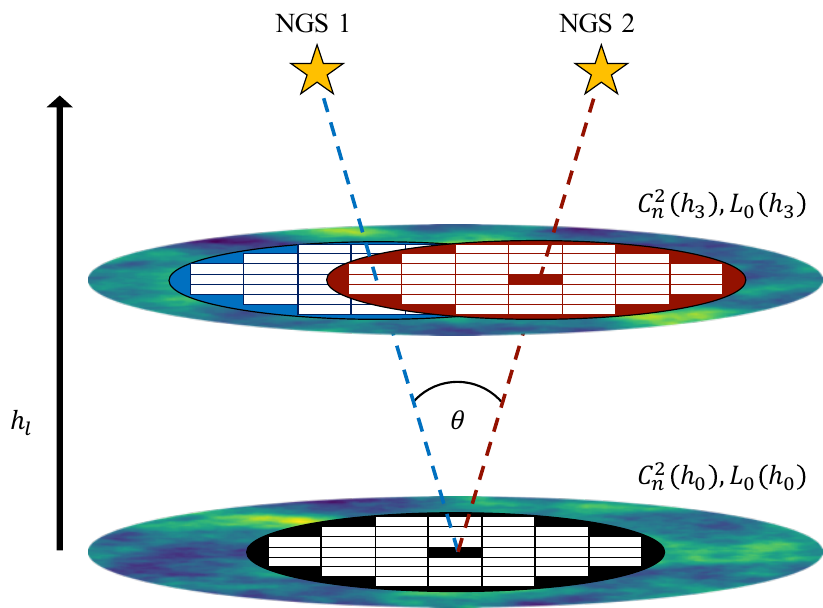}
    \caption{Sub-aperture optical paths of two $7\times7$ SHWFSs to 
    NGS~1 (black$\to$blue) and NGS~2 (black$\to$red). The two turbulent layers are at altitudes of 0 and $3D/7\theta$\,km.}
    \label{fig:layerHeights}
\end{figure}

The SLODAR technique has been outlined in previously published literature ({\color{dg}for a detailed description see 
\citealp{Butterley2006}}). We review the key concepts here.
The system requires sufficiently bright objects that can be used as Guide Stars (GSs) for
wavefront sensing. The optical phase of each GS can then be measured at the ground via independent SHWFSs. 
Fig.~\ref{fig:layerHeights} illustrates the sub-aperture optical paths of two optically aligned $7\times7$ SHWFSs that are observing independent 
natural guide stars (NGSs). 
The distance to NGS sub-aperture optical path intersection in Fig.~\ref{fig:layerHeights} is given by $h_{l}$, where

\begin{equation}
    h_{l} = \frac{ls_{\text{w}}}{\theta}.
	\label{eq:h}
\end{equation}
In equation~\ref{eq:h} $\theta$ is the angular separation between the NGSs. The distance between the centres of two 
adjacent subapertures is given by $s_{\text{w}}$. It should be noted that equation~\ref{eq:h} is only valid if 
the NGSs are separated in the FoV of the telescope 
by a position angle of 0, 90, 180 or $270^{\circ}$. Two altitudes of sub-aperture optical path intersection are shown in Fig.~\ref{fig:layerHeights}. 
The sub-aperture separation order is denoted $l$. The maximum distance of sub-aperture
optical path intersection, $h_{\textnormal{max}}$ ($h_{6}$ for the configuration shown in 
Fig.~\ref{fig:layerHeights}), is therefore $(D-s_{\text{w}})/\theta$, where $D$ is the diameter of the telescope. 
If an ELT-scale instrument and CANARY were observing the same NGS asterism, 
this implies that $h_{\textnormal{max}}$ would be larger for the ELT-scale instrument by a factor of $10.7$ (see Table~\ref{tab:params}). 
The ELT-scale instrument would also have an increased optical turbulence profile 
altitude-resolution. {$(h_{l+1}-h_{l})$ for CANARY and an 
ELT-scale instrument would be $0.6/\theta$ and $0.53/\theta$, respectively.

It has been shown that the SLODAR method can also utilise laser guide stars (LGSs; \citealp{Cortes2012}). 
{\color{dg}Due to the cone effect the distance to LGS sub-aperture optical path intersection, $a_{l}$, 
does not scale linearly with $l$. For optically aligned SHWFSs with LGS position angles of 0, 90, 180 or 
$270^{\circ}$}
\begin{equation}
    a_{l} = \frac{nls_{\text{w}}}{\theta n + ls_{\text{w}}},
	\label{eq:hlgs}
\end{equation}
where $n$ is the distance to the LGSs. Equation~\ref{eq:hlgs} assumes that $n$ is the same for both LGSs.

{\color{dg}The SLODAR technique is able to use SHWFS GS measurements to triangulate atmospheric turbulence strength as a function of 
altitude (see Section~\ref{sec:covMatrix}). These atmospheric aberrations have an outer scale, $L_{0}$.} 
Investigations indicate that the outer scale resides between 1-100\,m \citep{Ziad2004}. 
However, using the SLODAR technique the existing class of telescopes do not have the spatial
scale to construct an un-biased $L_{0}$ profile \citep{Ono2016}. It was shown by \cite{Guesalaga2017} that there is no significant difference 
in SLODAR data analysis when the outer scale is over roughly 3 times the diameter of the telescope pupil. 
For these reasons it is common practice for the existing class of telescopes to assume $L_{0}=25$\,m. 
ELT-scale telescopes might not be able to make this approximation. It is possible that the ELTs will 
have to measure $L_{0}(h)$.

\begin{figure}
    \centering
    \subfigure[Analytically generated CANARY covariance matrix for two $7\times7$ SHWFSs observing turbulent layers at altitudes of 
    0 and $3D/7 \theta$\,km. Both layers have $r_{0}=0.2$\,m and $L_{0}=25$\,m. \label{fig:covMat}]{\includegraphics{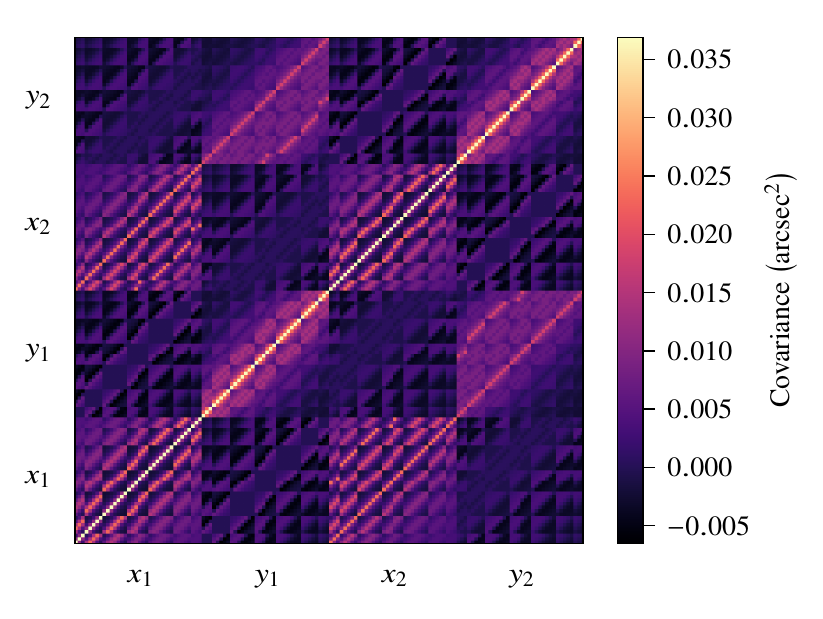}}
  
    \subfigure[The covariance map from (a). This shows the average covariance between $x_{1}x_{2}$ and $y_{1}y_{2}$ as a function of baseline. 
    \label{fig:covMap}]{\includegraphics{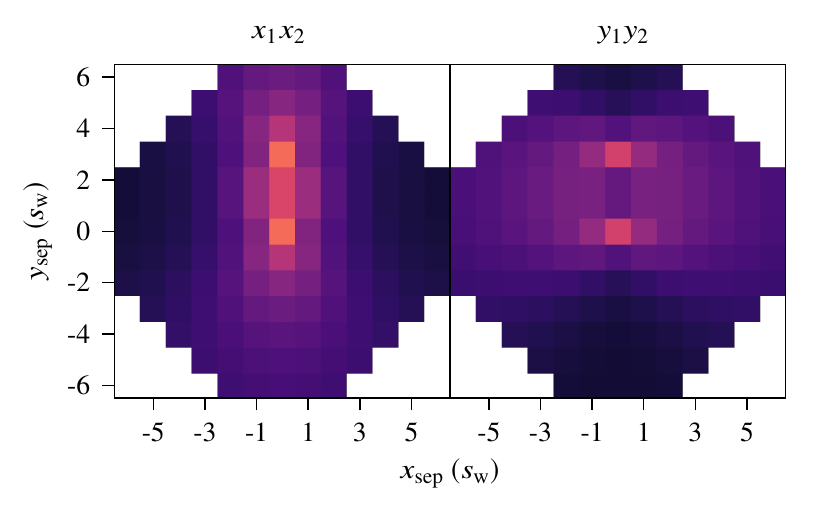}}
  
    \subfigure[The covariance map ROI from the data points in (b). In this example the length and width of the ROI is given by 
    $L=n_{\textnormal{d}}$ and $W=1$, respectively. \label{fig:covMapROI}]{\includegraphics{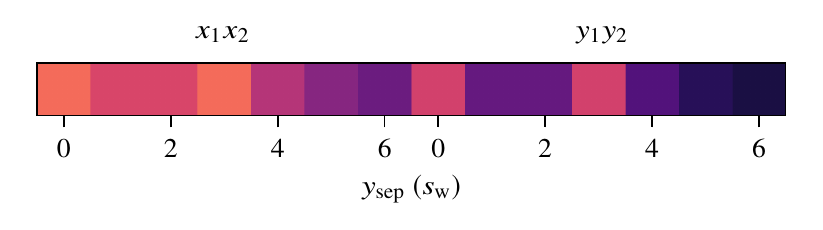}}
  
    \caption{Techniques for expressing the optical turbulence profile using the cross-covariance of SHWFS centroids. \label{fig:covs}}
\end{figure}

\subsection{Covariance matrix}
\label{sec:covMatrix}

The cross-covariance between all open-loop centroid positions 
(recorded over some time interval) expresses the optical turbulence profile as a covariance matrix.  
An analytically generated 2-NGS CANARY covariance matrix is shown in Fig.~\ref{fig:covMat}. 
It corresponds the optical turbulence profile illustrated in Fig.~\ref{fig:layerHeights} i.e. two 
layers at altitudes of 0 and $3D/7 \theta$\,km. 
Both turbulent layers are characterised by $L_{0}=25$\,m and $r_{0} = 0.2$\,m. $r_{0}$ 
is the Fried parameter. Orthogonal measurements 
from NGS~1 and NGS~2 are given by $x_{1}$, $y_{1}$ and
$x_{2}$, $y_{2}$, respectively. It can be seen in Fig.~\ref{fig:covMat} that the lowest signal-to-noise 
(SNR) cross-covariance measurements are between orthogonal axes. 
The cross-covariance between sub-apertures 
from the same SHWFS i.e. the auto-covariance, gives a measure of integrated turbulence strength. 
The strongest response to the vertical structure of the optical turbulence profile is between equivalent planes 
of independent SHWFSs e.g. cov$(x_{1}, x_{2})$ and cov$(y_{1}, y_{2})$.

\subsection{Covariance map}
\label{sec:covMap}

We define baseline as the two-dimensional sub-aperture separation between two optically 
aligned SHWFSs at $l=0$. The baseline between SHWFSs 
is characterised by sub-aperture ordering and is therefore independent of any misalignment. The baseline in $x$ and $y$ is given by 
$x_{\text{sep}}$ and $y_{\text{sep}}$, respectively.
The cross-covariance between equivalent planes of independent SHWFSs ($x_{1}x_{2}$ and $y_{1}y_{2}$ in Fig.~\ref{fig:covMat}) 
can be averaged as a function of baseline. The resultant array is known as a covariance map. 
The covariance map from Fig.~\ref{fig:covMat} is shown in Fig.~\ref{fig:covMap}. The covariance map contains all the high SNR
information for the vertical structure of the optical turbulence profile. 
If a system is characterised by random noise then averaging baselines increases the SNR of the optical turbulence profile. 
Any two GSs in the FoV of the telescope are separated by a position angle, $\gamma$. 
In Fig.~\ref{fig:layerHeights} $\gamma=0$\,rad and as shown in Fig.~\ref{fig:covMap}, 
the optical turbulence profile is primarily projected over sub-aperture separations that are determined by $\gamma$.

\subsection{Covariance map region of interest}
\label{sec:covMapROI}

Measurements within the
covariance map can be extracted along the vector projected by $\gamma$ - a covariance map Region of Interest (ROI). 
This requires $\gamma$ to be known with respect to the geometry of the SHWFS lenslet array.
The covariance map ROI in Fig.~\ref{fig:covMapROI} is 
taken from the centre of $x_{1}x_{2}$ and $y_{1}y_{2}$ in Fig.~\ref{fig:covMap}, and along positive $y_{\text{sep}}$. In reality the optical turbulence profile
may not be projected across exact baselines (as it is in Fig.~\ref{fig:covMapROI}). 
The ROI can compensate for this by encapsulating a
larger extent of the map i.e. its length and width can be increased. In this study the 
length and width of the ROI are denoted by $L$ and $W$, respectively. Both $L$ and $W$ are in 
units of the distance between two adjacent sub-apertures, $s_{\text{w}}$. Fig.~\ref{fig:covMapROI} has $L=n_{\textnormal{d}}$ and $W=1$, where 
$n_{\textnormal{d}}$ is the number of SHWFS sub-apertures in one dimension. $L>n_{\text{d}}$ for the configuration in 
Fig.~\ref{fig:covMapROI} implies that the ROI includes negative $y_{\text{sep}}$ data points e.g. $L=(n_{\text{d}}+1)$ is the 
same ROI in Fig.~\ref{fig:covMapROI} but with the inclusion of $x_{\text{sep}},y_{\text{sep}}=(0,-1)$ in both $x_{1}x_{2}$ 
and $y_{1}y_{2}$ from Fig.~\ref{fig:covMap}. An ROI of $L>n_{\text{d}}$ can be thought of as extending 
the ROI to data points that are projected along $\gamma+\pi$.
It should be noted that $W=1$ reduces $L_{0}(h)$ information. There are current as well as forthcoming AO 
systems that utilise more than two GSs. This assists 
wide-field AO as it introduces the capability of analysing the optical turbulence profile 
at multiple altitude resolutions. The covariance map ROI studies the 
optical turbulence profile in a multi-GS system by stacking the ROI from each GS combination into a single array.

The measured covariance map ROI does not have to come from calculating the covariance matrix and 
then the covariance map i.e. the steps from Section~\ref{sec:covMatrix} to \ref{sec:covMap}. We have developed an algorithm that calculates the 
covariance map ROI directly from centroid measurements. This process utilises a look-up table that lists sub-aperture combinations 
as a function of baseline. The algorithm uses $\gamma$ with this look-up table to calculate the average cross-covariance 
at the required baselines.

\section{Optical turbulence profiling}
\label{sec:levMar}

It is possible to develop an algorithm for analytically generating the covariance between sub-apertures. Analytical 
covariance for each turbulent layer depends on $r_{0}$, $L_{0}$, $h$, sub-aperture separation and optical misregistrations. It should 
again be noted that baseline is independent of SHWFS misalignments.
By analytically generating sub-aperture covariance we are able to iteratively fit an 
analytical model to SHWFS cross-covariance 
measurements. This allows us to measure the optical turbulence profile - the idea of the learn stage 
within learn \& apply MOAO tomography \citep{Vidal2010}. For more details on 
the model fitting procedure and the open-source software package that we have developed see Section~\ref{sec:capt}.
The reader is directed to \citealp{Martin2016} for a complete description of the mathematical procedure 
that we use for analytically generating sub-aperture covariance.

The approximation we use for analytically generating sub-aperture covariance has been shown to 
increase in accuracy the further the projected sub-apertures are separated. At sub-aperture separations of zero 
the analytical model has an overestimation of $\sim15\%$ \citep{Martin2012}. Furthermore, measured covariance at a 
sub-aperture separation of zero is susceptible to a low SNR. For these reasons the model fitting procedure we use sets 
the measured cross-covariance between all sub-apertures with baseline $x_{\text{sep}},y_{\text{sep}}=(0,0)$ to zero.
It also sets analytically generated covariance at all $x_{\text{sep}},y_{\text{sep}}=(0,0)$ baselines to zero.

\subsection{Subtracting ground-layer isoplanatic turbulence}
\label{sec:l3s}

{\color{dg}The strong ground-layer is usually slow-moving and therefore its cross-covariance function can differ from the 
analytical model.} This makes it difficult 
to accurately measure the optical turbulence profile at non-zero altitudes. Learn 3 step (L3S; \citealp{Martin2016}) is a fitting technique 
used to better detail the optical turbulence profile at non-zero altitudes. It does this by initially subtracting ground-layer isoplanatic turbulence. 
Subtracting centroid ground-layer isoplanatic turbulence requires calculating the mean centroid location in $x$ and $y$ for every frame at each 
sub-aperture baseline. All centroids from every SHWFS then have their respective average at each frame subtracted, 
removing common-motion at each sub-aperture baseline. In an optically aligned system this common-motion subtraction mitigates ground-layer turbulence. 
It also simultaneously removes internal 
vibration artefacts - remnants of wind-shake and telescope tracking errors, seen as a linear addition to 
cross-covariance measurements with independent values at $xx$ and $yy$.
As this treatment is linear the transformation matrix, {\sffamily$\textbf{\text{T}}$}, used to perform common-motion subtraction on the SHWFS 
centroids can be directly applied to an analytically generated covariance matrix. Having {\sffamily$\textbf{\text{A}}$} and {\sffamily$\textbf{\text{B}}$} represent an 
analytically generated covariance matrix before and after ground-layer mitigation, respectively, implies that
{\sffamily{\begin{equation}
    \textbf{\text{B}} = \textbf{\text{T}} \cdot \textbf{\text{A}} \cdot \textbf{\text{T}}^{\textnormal{T}}.
	\label{eq:l3s}
\end{equation}}} 
An analytical covariance map ROI with subtracted ground-layer isoplanatic turbulence can be calculated by averaging the specific 
baselines within {\sffamily$\textbf{\text{B}}$} (see Section~\ref{sec:covMapROI}).

\subsection{Covariance parameterisation of optical turbulence and SHWFS misalignments}
\label{sec:capt}

In this study we use a fitting technique similar to L3S. This decision was based on the fact that L3S 
has been shown to be robust and to minimise tomographic error \citep{Martin2012}. 
We use the Levenberg Marquardt algorithm (LMA) to fit an analytical model to SHWFS cross-covariance measurements. 
{\color{dg}The software package we have developed is capable of supervising an AO system by using 
AO telemetry to measure the optical turbulence profile and SHWFS misalignments in real-time. The user chooses 
how many layers to fit and the altitudes at which these layers are fitted. Multiple GS combinations can also be 
fitted to simultaneously.} This software package is now open-source\footnote{https://github.com/douglas-laidlaw/CAPT} 
(example test cases are included in the open-source package). The software is written in Python and it uses 
the numpy (\citealp{Vanderwalt2011}) and scipy (\citealp{Jones}) libraries. 
We refer to the AO system supervisor that 
we have developed as covariance parameterisation of optical turbulence and SHWFS misalignments (CAPT). 
The three steps of CAPT for a NGS system are as follows.
\begin{description}
    \item 1. Using the transformation matrix, {\sffamily$\textbf{\text{T}}$}, remove centroid common-motion and calculate the chosen cross-covariance array 
    e.g. covariance matrix or map ROI. The LMA fits to measured cross-covariance with an analytically generated covariance 
    array that is ground-layer mitigated. The fit is performed by iteratively adjusting $C_{n}^{2}(h)$. 
    It is assumed that $L_{0}(h)=25$\,m. 
    \item 2. $C_{n}^{2}(h>0)$ from 1 is used to analytically generate a covariance array that dissociates ground-only turbulence from 
    the complete cross-covariance array (the measured cross-covariance array with no ground-layer mitigation). The LMA 
    fits analytically generated covariance to the measured ground-only cross-covariance array. 
    The fit is performed by iteratively adjusting $C_{n}^{2}(0)$, $L_{0}(0)$ and vibration artefacts.
    \item 3. Using the parameters recovered from 1 and 2, the LMA fits analytically generated covariance to the complete 
    cross-covariance array by iteratively adjusting SHWFS shift and rotation misalignments.
\end{description}
As mentioned previously the fitting technique we have adopted closely resembles L3S. 
The differences between CAPT and L3S are listed below.
\begin{itemize}
    \item The first and second steps of L3S remove tip-tilt in both the measured and 
    analytically generated cross-covariance arrays. We found that this did not improve the results and so tip-tilt removal was 
    not included in the first and second steps of CAPT (CAPT~1 and CAPT~2, respectively). Removing tip-tilt also requires an 
    additonal matrix multiplication. Not including this operation therefore improved the efficiency of the system.
    \item The first step of L3S fits $C_{n}^{2}(h>0)$ and $L_{0}(h>0)$. Although CAPT is capable of fitting an outer scale profile, 
    we do not fit $L_{0}(h)$ for reasons discussed in 
    Section~\ref{sec:slopeCov}. We fit $C_{n}^{2}(0)$ during CAPT~1 to help account for possible SHWFS misalignments. 
    However, the measurement of $C_{n}^{2}(0)$ is taken from CAPT~2.
    \item The third step of L3S fits vibrational artefacts. We are able to complete this in CAPT~2 as the system is not 
    tip-tilt subtracted. L3S also fits $xx$, $yy$ and $xy$ vibration artefacts. CAPT does not fit $xy$ vibrational artefacts as 
    it assumes orthogonal centroids are decorrelated. 
    \item The third step of L3S fits SHWFS magnification. We do not fit this parameter during the third step of CAPT (CAPT~3) 
    as this study concentrates on the parameterisation of SHWFS misalignments.
\end{itemize}

The outer scale measurement in CAPT~2 is not considered physical. It is fitted to acount for the liklihood 
that the analytical model will differ from the cross-covariance function of the slow-moving ground-layer. 
During CAPT~1 and CAPT~2, the covariance map ROI has $L=n_{\textnormal{d}}$ and $W=1$. In CAPT~3 
$L=(n_{\textnormal{d}}+1)$ and $W=3$. The ROI in 
CAPT~3 is increased so that it has enough spatial information to detect SHWFS misalignments. These are the dimensions of the covariance map ROI 
unless stated otherwise. 
The noise-floor for $C_{n}^{2}(h)dh$ is set at $10^{-16}$\,m$^{1/3}$ i.e. measurements of $C_{n}^{2}(h)dh < 10^{-16}$\,m$^{1/3}$
are made equal to $10^{-16}$\,m$^{1/3}$.

The uplink of LGSs through atmospheric turbulence results in the loss of tip-tilt information. 
This causes each SHWFS to have an independent tip-tilt term after CAPT~1 ground-layer mitigation. The result of this is a 
cross-covariance discontinuity within each $xx$ and $yy$ measurement. During LGS optical turbulence profiling, CAPT~1 fits linear additions to 
each $xx$ and $yy$ term to account for this discontinuity. In CAPT~2 this fitted parameter is used to help 
dissociate the ground-layer. It is also included during CAPT~3.

The vertical structure of the optical turbulence profile is dependent on sub-aperture separation. An unknown 
SHWFS misalignment will therefore directly impact the accuracy of CAPT. SHWFS shift and rotation 
misalignments are fitted during CAPT~3 however, the optical turbulence profile is recovered during CAPT~1 and CAPT~2. This implies that if 
an unknown misalignment exists between SHWFSs, the optical turbulence profile will be imprecisely recovered which will lead to 
CAPT~3 being unable to recover systematic uncertainties. 
We found that a solution to this problem is to iterate through CAPT several times i.e. run CAPT~1 and CAPT~2, take 
misalignment measurements from CAPT~3, put them back into CAPT and repeat (see Section~\ref{sec:meas_misalign}). 
The idea is that after a number of iterations CAPT will converge on an accurate measure of 
both $C_{n}^{2}(h)$ and the alignment between the SHWFSs. At the first iteration every layer in the fitted model has a starting 
point of $r_{0}=0.2$\,m and $L_{0}=25$\,m. Vibrational artefacts and SHWFS misalignments have a starting point of zero. As we iterate 
through CAPT a number of times, the starting point of each fitted parameter is equal to its measurement from the previous iteration.

\section{Results}
\label{sec:results}

\subsection{SHWFS misalignments and the degradation of optical turbulence profiling}
\label{sec:misalignments}

In a physical AO system it is unrealistic to assume perfect optical alignment. It is therefore important to quantify the degradation of
covariance matrix and map ROI optical turbulence profiling in the presence of SHWFS misalignments. 
For CANARY, AOF and ELT-scale configurations, a 2-NGS 
covariance matrix was analytically generated for 11 evenly-spaced SHWFS rotational misalignments. The maximum offset 
was $5^{\circ}$. The target covariance matrix (the covariance matrix that would be fitted to), $M_{\textnormal{T}}$, 
for each AO system was analytically generated with no noise. 
This allowed for perfect convergence of centroid cross-covariance measurements to be assumed. 
Each AO system had $M_{\textnormal{T}}$ generated with $\theta$ such that $h_{\textnormal{max}} = 24$\,km and $\gamma=0$\,rad. 
Median atmospheric parameters documented by the European southern observatory (ESO; \citealp{eso2015}) 
were used to parameterise the optical turbulence profile. Integrated $r_{0}$ was $0.1$\,m. For consistency between each AO system we chose to fit 
seven evenly-spaced layers from 0 to 24\,km. The total number of measured layers is given by $N_{\text{L}}$ i.e. $N_{\text{L}}=7$. 
The measured altitudes are given by $h^{\text{m}}_{i}$, where $i$ denotes the layer number e.g. $h_{3}^{\text{m}}=8$\,km.  
The $C_{n}^{2}(h)$ profile for each $M_{\textnormal{T}}$ was calculated by binning the 35-layer ESO profile 
into these 7 evenly-spaced layers i.e. if the evenly-spaced layers have an altitude width $b_{\text w}$, at altitude $h_{i}^{\text m}$ 
the 35-layer ESO profile is integrated between $h^{\text m}_{i}-b_{\text w}/2$ and $h^{\text m}_{i}+b_{\text w}/2$. 
$L_{0}$ for each of the 7 layers in $M_{\text{T}}$ was 25\,m. The parameterised and fitted turbulent altitudes were made equal 
to compensate for each telescope yeilding a unique 
optical turbulence profile resolution (due to their specified SHWFS sub-aperture configuration in Table~\ref{tab:params}) i.e. 
for zero SHWFS misalignments in each AO system, both covariance matrix and map ROI fitting would recover the exact optical turbulence profile.

Covariance matrix CAPT was performed at each rotationally offset $M_{T}$. 
The covariance map ROI was calculated from each $M_{T}$ and covariance map ROI CAPT was also performed.
For covariance matrix and map ROI fitting, CAPT assumed a rotational offset of zero. This 
allowed for the degradation of the measured optical turbulence profile to be monitored as a 
known rotational offset was introduced. The mean deviation, $F_{\textnormal{md}}$, between the fitted and parameterised 
profile was used to quantify the results. To account for the wide-range of turbulence strengths $F_{\textnormal{md}}$ 
was performed in logarithmic space such that
\begin{equation}
    \label{eq:fmd}
    \begin{aligned}
        F_{\textnormal{md}} = \frac{1}{N_{\text{L}}} \sum_{i=1}^{N_{\text{L}}} \Big|\log_{10} \Big(C_{n}^{2}(h_{i}^{\text{m}})^{\text{m}} \Big/ C_{n}^{2}(h_{i}^{\text{m}})^{\text{r}}\Big)\Big|.
    \end{aligned}
\end{equation}
Put simply, $N_{\text{L}} \cdot F_{\textnormal{md}}$ is the total order of magnitude difference 
between the measured and reference optical turbulence profiles ($C_{n}^{2}(h^{\text{m}})^{\text{m}}$ and 
$C_{n}^{2}(h^{\text{m}})^{\text{r}}$, respectively).
In equation~\ref{eq:fmd} $C_{n}^{2}(h^{\text{m}})^{\text{r}}$ is the binned 35-layer ESO profile. 
The $F_{\textnormal{md}}$ results for an increasing SHWFS rotational misalignment are shown in Fig.~\ref{fig:rotOffset}. 
For the CANARY system, the steep increase in covariance matrix $F_{\textnormal{md}}$ is seen as the optical 
turbulence profiling procedure begins to incorrectly detect zero turbulence at 24\,km. The covariance map ROI does not suffer 
from this. AOF and ELT-scale 
$F_{\textnormal{md}}$ results favour the covariance matrix. The most likely reason for the covariance matrix being more robust 
is that its number of covariance measurements is inversely proportional to sub-aperture separation. This implies that 
small sub-aperture separations carry more weight during CAPT. Sub-apertures separated by the largest distance will be most affected 
by a rotational misalignment and so this weighting is a benefit of covariance matrix fitting. 
We chose to not weight the covariance map ROI accordingly as in a real-world system this might 
amplify noise. However, appropriate covariance map ROI weighting should be the subject of a future investigation. 
In Fig.~\ref{fig:rotOffset} there is no significance to ELT-scale covariance matrix and map ROI 
fitting having similar $F_{\textnormal{md}}$ values at a rotational offset of $5^{\circ}$. At $F_{\textnormal{md}}\simeq0.65$, $C_{n}^{2}(h_{i}^{\text{m}})^{\text{m}}$ 
is entirely unrepresentative of $C_{n}^{2}(h_{i}^{\text{m}})^{\text{r}}$.
Furthermore, the plots shown in Fig.~\ref{fig:rotOffset} are specific to the parameterised optical 
turbulence profile i.e. $F_{\textnormal{md}}$ results are 
dependent on $C_{n}^{2}(h)^{\textnormal{r}}$, $\theta$ and $\gamma$. $F_{\textnormal{md}}$ results 
will also be dependent on the AO system e.g. $s_{\text{w}}$.
The purpose of Fig.~\ref{fig:rotOffset} is to outline the scale of the problem 
when there is an unknown SHWFS rotational misalignment.

The same investigation was performed but for lateral shifts in SHWFS alignment i.e. a shift misalignment 
equal in both $x$ and $y$. SHWFSs are conjugate to the pupil of the telescope (see Fig.~\ref{fig:layerHeights}) 
and so the lateral shift was set as a function of telescope diameter, $D$. 
Having the lateral shift as a function of $D$ meant that the scale of the misalignment was proportional for each AO system. 
The maximum shift was set at an offset of 5\% of $D$. The results are shown 
in Fig.~\ref{fig:shiftOffset}. It is clear that there is a resemblance between Fig.~\ref{fig:rotOffset} 
and Fig.~\ref{fig:shiftOffset}. The reason for this is that in both cases sub-aperture separation in
$x$ and $y$ is dependent on $D$. This is also why 
the general trend is for larger telescopes to experience higher 
$F_{\textnormal{md}}$ values. It should again be noted that these $F_{\textnormal{md}}$ results are 
dependent on both the AO system and the parameterised optical 
turbulence profile.

\begin{figure}
	\includegraphics{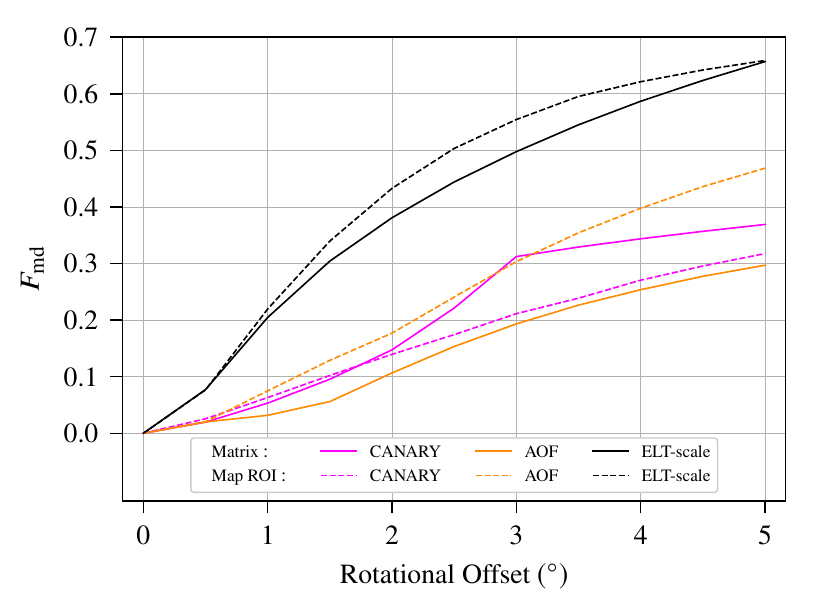}
    \caption{The degradation in the accuracy of matrix and map ROI CAPT optical turbulence profiling as a SHWFS rotation misalignment is introduced.}
    \label{fig:rotOffset}
\end{figure}
\begin{figure}
	\includegraphics{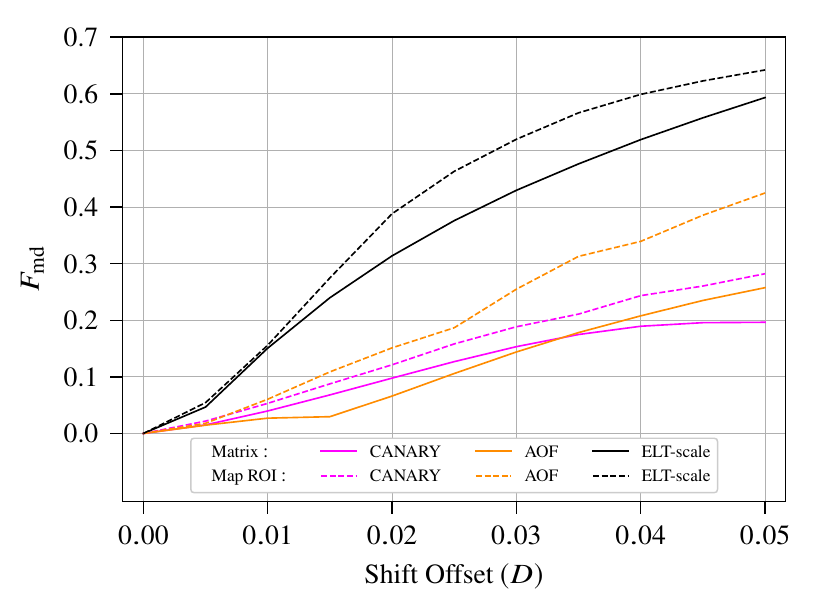}
    \caption{The degradation in the accuracy of matrix and map ROI CAPT optical turbulence profiling as a SHWFS shift misalignment is introduced.}
    \label{fig:shiftOffset}
\end{figure}

To illustrate the degradation of the optical turbulence profile with increasing $F_{\textnormal{md}}$ values, 
$C_{n}^{2}(h_{i}^{\text{m}})^{\text{m}}$ is shown in Fig.~\ref{fig:fmd_comp} for ELT-scale covariance matrix fitting 
at SHWFS shifts of 0.00, 0.02 and $0.05D$. These shifts correspond 
to $F_{\textnormal{md}}$ values of roughly 0.0, 0.3 and 0.6 (see Fig.~\ref{fig:shiftOffset}). They also correspond to $r_{0}$ seeing 
values of 0.10, 0.10 and 0.11\,m. This indicates why $F_{\textnormal{md}}$ is the preferred metric for determining the accuracy 
of $C_{n}^{2}(h^{\text{m}})^{\text{m}}$. 
$C_{n}^{2}(h^{\text{m}})^{\text{m}}$ at a shift misalignment of $0.00D$ ($F_{\textnormal{md}}=0.0$) in Fig.~\ref{fig:fmd_comp} is 
synonymous with $C_{n}^{2}(h^{\text{m}})^{\text{r}}$. At a shift misalignment of $0.05D$ 
the form of the reference optical turbulence profile has been completely lost. At a shift misalignment of $0.02D$ ($F_{\textnormal{md}}\simeq0.3$)
$C_{n}^{2}(h^{\text{m}})^{\text{m}}$ is beginning to significantly deviate from $C_{n}^{2}(h^{\text{m}})^{\text{r}}$ 
- especially at 20 and $24$\,km. CANARY and AOF reach this level of deviation when they are subject to either an unknown rotational or 
shift misalignment of around $5^{\circ}$ and $0.05D$, respectively. For an ELT-scale optical turbulence 
profiling system to achieve this level of accuracy these values must be below $1.5^{\circ}$ and $0.02D$. 
If an ELT-scale system can be optically aligned to this level of accuracy $C_{n}^{2}(h^{\text{m}})^{\text{m}}$ will not 
significantly deviate from the real optical turbulence profile. Alternatively, CAPT can measure and thus 
compensate for SHWFS misalignments (see Section~\ref{sec:meas_misalign}).

\subsection{Measuring SHWFS misalignments}
\label{sec:meas_misalign}

SHWFS misalignments are a concern for 
reliably measuring the optical turbulence profile during CAPT~1 and CAPT~2, and are therefore problematic when estimating 
systematic misregistrations during 
CAPT~3. Using the same CANARY-scale configuration from Sec~\ref{sec:misalignments}, CAPT was performed for 200 datasets that had SHWFS misalignments 
in both rotation and lateral shift. The value of these misalignments were randomly generated from a uniform distribution. Rotation and 
shift misalignments had a range of $-5^{\circ}$ to $5^{\circ}$ and $-0.05D$ to $0.05D$, respectively. 
Covariance matrix and map ROI CAPT was repeated 15 times for each dataset and $F_{\textnormal{md}}$ was 
calculated at each iteration. For CAPT~1 and CAPT~2 the ROI had $W=1$ and $L=n_{\text{d}}$. The ROI in CAPT~3 had 
$W=3$ and $L=(n_{\text{d}}+1)$ so that there was enough spatial information to fit SHWFS misalignments. 
$F_{\textnormal{md}}\to0$ implies that CAPT~1 and CAPT~2 are 
successfully converging on $C_{n}^{2}(h^{\text{m}})^{\text{r}}$, and that CAPT~3 is accurately 
measuring the misalignment between the SHWFSs. $F_{\textnormal{md}}$ will not equal 0 unless both of these conditions are satisfied. 
The results in Fig.~\ref{fig:measOffset1} show that both covariance matrix and map ROI fitting 
converge on the solution for all SHWFS misalignments. 

The covariance matrix has a lower $F_{\textnormal{md}}$ value after the first CAPT 
iteration for reasons discussed in Section~\ref{sec:misalignments}. On average both the covariance matrix and map ROI 
accurately measure the SHWFS misalignment after the first iteration. The outliers imply that there are particular SHWFS misalignments that require 
a number of CAPT iterations. 15 CAPT iterations guarantees statistical convergence. The remainder of Section~\ref{sec:results}
shall therefore operate a system that iterates through CAPT 15 times. During the operation of a real-world system the idea 
is that SHWFS misalignments will not have to be fitted every time the optical turbulence profile is measured. 
SHWFS misalignments would be logged periodically so that only CAPT~1 and CAPT~2 have to be performed.

\begin{figure}
	\includegraphics{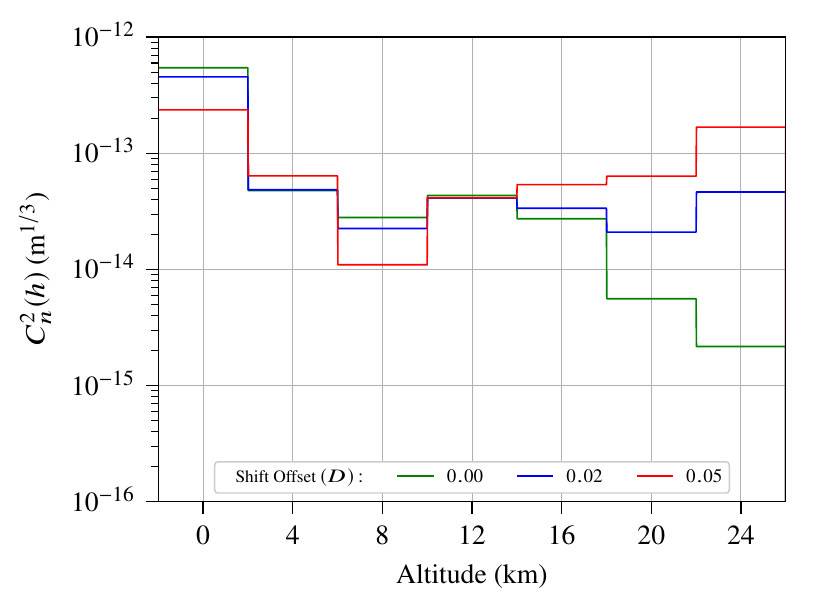}
    \caption{The degradation of ELT-scale covariance matrix $C_{n}^{2}(h^{\text{m}})^{\text{m}}$ as a lateral shift misalignment 
    is introduced. Shift offsets of 0.00, 0.02 and $0.05D$ correspond to $F_{\textnormal{md}}$ values of $0.0$, $\sim0.3$ and $\sim0.6$ (see Fig.~\ref{fig:shiftOffset}).}
    \label{fig:fmd_comp}
\end{figure}

\subsection{Simulated open-loop AO telemetry optical turbulence profiling}
\label{sec:simResults}

This section investigates the quality of optical turbulence profiling that can be achieved by using 
CAPT with AO telemetry. 
NGS SHWFS open-loop centroids were generated for
the CANARY configuration in SOAPY\footnote{https://github.com/AOtools/soapy}: a Monte Carlo AO simulation package (\citealp{Reeves2016}). 
The median results within 
the ESO documentation were used to parameterise the simulated 35-layer profile. Integrated $r_{0}$ was $0.1$\,m 
and $L_{0}$ at each of the 35 layers was 25\,m. Four optically aligned $7\times7$ 
SHWFSs observing independent NGSs at zenith - all of 10$^{\textnormal{th}}$ apparent magnitude in the V-band - 
were simulated in a square layout with $h_{\textnormal{max}} = 24$\,km. The 6 NGS combinations had $\gamma$ values of 
0, 45, 90, 90, 135 and $180^{\circ}$. On-sky observations do not have such well-ordered NGS asterisms with equal apparent magnitudes. 
However, the simulated model is an appropriate approximation. All SHWFSs were simulated to measure $500$\,nm wavelengths and 
each sub-aperture recorded 10,000 measurements. 
These were calculated by the centre-of-gravity at each frame and were subject to both shot and read-out 
noise (approximated to 1 electron per pixel). The throughput of the system was 30\% and the frame rate was 150\,Hz. 
The simulation was repeated ten times so that the mean along with its standard error could be calculated.

\begin{figure}
	\includegraphics{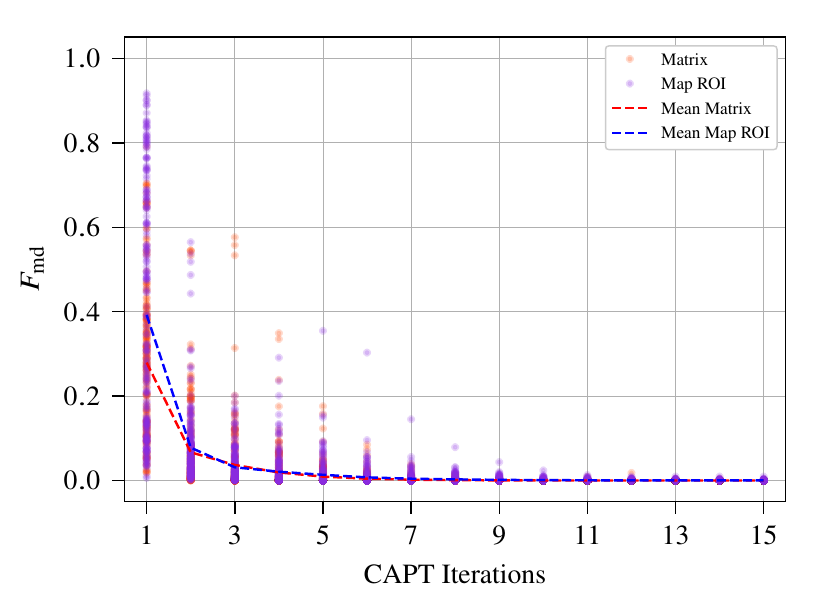}
    \caption{Covariance matrix and map ROI $F_{\textnormal{md}}$ for 200 datasets over 15 CAPT
    iterations. The results are shown for a CANARY-scale 2-NGS configuration where $N_{\text{L}}=7$. The SHWFSs in $M_{\text{T}}$ 
    were randomly misaligned in both rotation and lateral shift.}
    \label{fig:measOffset1}
\end{figure}

Covariance matrix and map ROI CAPT fitting procedures were performed on their respective cross-covariance arrays. 
The 6 SHWFS combinations were fitted to simultaneously. Each algorithm fitted seven evenly-spaced 
layers from 0 to 24\,km. Vibration artefacts and SHWFS misalignments were not included in the simulation and were therefore 
measured to be negligible. The 35-layer ESO profile was binned to the fitted altitudes so that the 
measured optical turbulence profile could be compared to a reference. 
$F_{\textnormal{md}}$ was used to quantify profiling accuracy. The results are given in Table~\ref{tab:simResults}. 
The mean $C_{n}^{2}(h^{\text{m}})^{\text{m}}$ profiles are shown in Fig.~\ref{fig:ngs_barPlot} alongside 
$C_{n}^{2}(h^{\text{m}})^{\text{r}}$. In Fig.~\ref{fig:ngs_barPlot} the most noticeable difference 
between covariance matrix and map ROI optical turbulence profiling is at 20 and 24\,km. 
As there are a reduced number of optical turbulence profile 
measurements at higher altitudes, the most probable cause for the covariance map 
ROI being more accurate is that it only considers the highest SNR measurements.
If only the first 5 layers are considered (0$\to$16\,km; see Table~\ref{tab:simResults}) covariance matrix fitting is more 
accurate due to its ground-layer measurement. This is likely caused by 
auto-covariance measurements constraining the model during CAPT~2. However, by studying 
Fig.~\ref{fig:ngs_barPlot} there is little difference in the 
$F_{\textnormal{md}}$ results when analysing the layers fitted from 0 to 16\,km.

The NGS study was repeated for sodium LGSs. All system 
parameters were the same except for the apparent magnitude of each LGS in the V-band. This was set to 8. Each 
LGS was side-launched and focussed at an altitude of $90$\,km. The effects of LGS elongation and fratricide were not included 
in the simulation however, on a 4.2\,m telescope these effects have little impact on the results. The SHWFSs were simulated to measure $589$\,nm 
wavelengths - the wavelength of sodium D-lines. Fig.~\ref{fig:ngs_barPlot} is repeated for the LGS 
results in Fig.~\ref{fig:lgs_barPlot}.  Due to the cone effect 
the maximum altitude of sub-aperture optical path intersection for this particular LGS asterism is $\sim19$\,km. Therefore, 
a layer was not fitted at $24$\,km as the turbulence at this altitude is unsensed. The layer fitted at $20$\,km is not 
included in the analysis of the measured optical turbulence profile as its bin is centred above the maximum altitude of sub-aperture 
optical path intersection. Measured vibration artefacts and SHWFS misalignments were again negligible. The LGS $F_{\textnormal{md}}$ 
results are summarised in Table~\ref{tab:simResults}. 
There is little difference between covariance matrix and map ROI LGS $F_{\textnormal{md}}$.

\begin{table}
    \centering
    \caption{Covariance matrix and map ROI optical turbulence profiling results from fitting to simulated 
    NGS and LGS CANARY cross-covariance arrays.}
    \begin{tabular}{|l|c|c|c|}
    \hline
            & \multicolumn{3}{c|}{$F_{\textnormal{md}}$}                                                                                \\ \hline
            & \multicolumn{1}{l|}{NGS, 0$\to$24\,km} & \multicolumn{1}{l|}{NGS, 0$\to$16\,km} & \multicolumn{1}{l|}{LGS, 0$\to$16\,km} \\ \hline
    Matrix  & 0.19 $\pm$ 0.03                    & 0.02 $\pm$ 0.00                    & 0.03 $\pm$ 0.00                    \\ \hline
    Map ROI & 0.13 $\pm$ 0.03                    & 0.03 $\pm$ 0.00                    & 0.05 $\pm$ 0.00                    \\ \hline
    \label{tab:simResults}
\end{tabular}
\end{table}
\begin{figure}
	\includegraphics{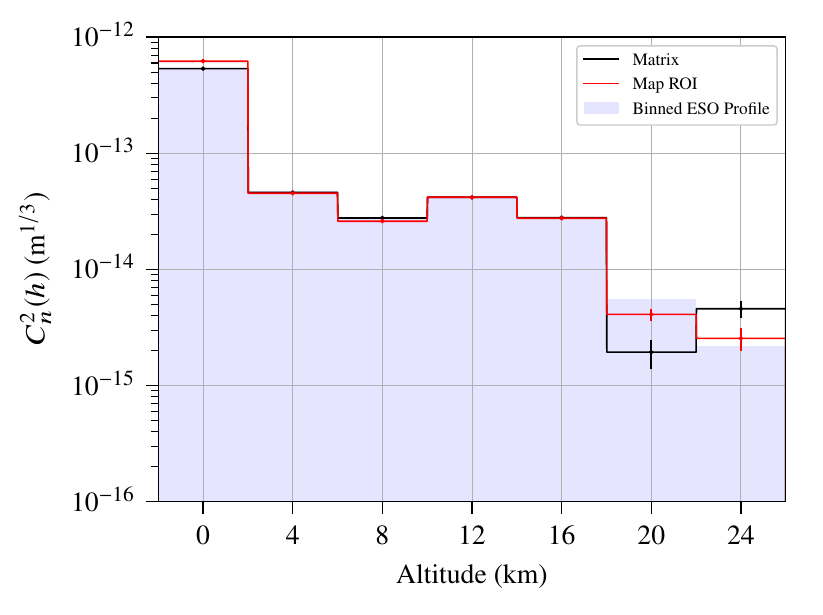}
    \caption{Covariance matrix and map ROI optical turbulence profiles from fitting to simulated 
    NGS CANARY cross-covariance arrays. The mean $C_{n}^{2}(h^{\text{m}})^{\text{m}}$ profiles are shown along with $C_{n}^{2}(h^{\text{m}})^{\text{r}}$.}
    \label{fig:ngs_barPlot}
\end{figure}
\begin{figure}
	\includegraphics{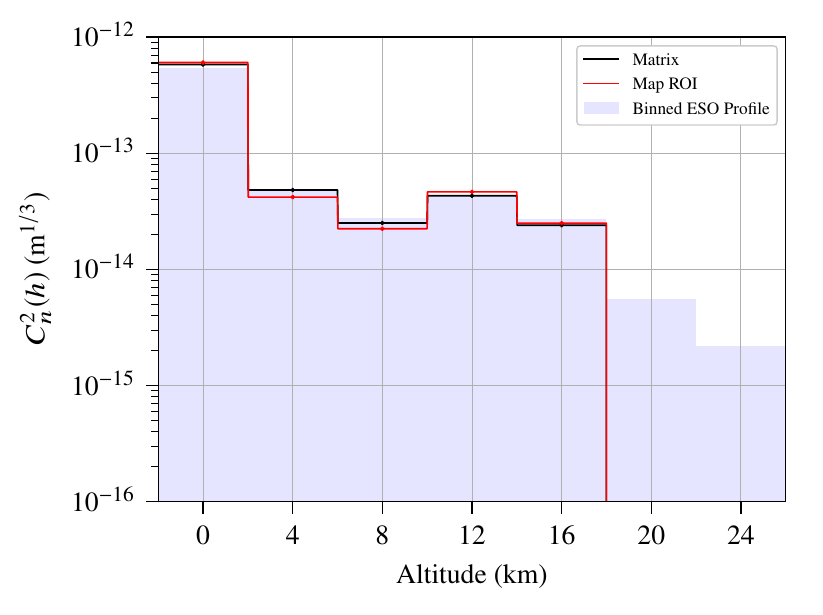}
    \caption{Covariance matrix and map ROI optical turbulence profiles from fitting to simulated 
    LGS CANARY cross-covariance arrays. The mean $C_{n}^{2}(h^{\text{m}})^{\text{m}}$ profiles are shown along with $C_{n}^{2}(h^{\text{m}})^{\text{r}}$.}
    \label{fig:lgs_barPlot}
\end{figure}

To investigate whether the covariance map ROI has an optimal size for recovering the optical turbulence profile, 
CAPT was repeated on the simulated NGS centroids but for all possible ROI widths and lengths. For consistency - and 
because it has been shown to be sufficient in Section~\ref{sec:meas_misalign} - the ROI during CAPT~3 kept values of 
$W=3$ and $L=(n_{\textnormal{d}}+1)$. 
$F_{\textnormal{md}}$ is plotted as a function of $W$ and $L$ in Fig.~\ref{fig:esoWL}. 
The trend in Fig.~\ref{fig:esoWL} indicates that profiling accuracy degrades 
as the ROI considers a greater number of baselines. The highest SNR optical turbulence profile 
information is along the NGS position angle, $\gamma$. Fig.~\ref{fig:esoWL} demonstrates that the system does not 
benefit from including baselines outside of $W=1$. The lowest and 
highest values within Fig.~\ref{fig:esoWL} are found at $L,W=(11,1)$ and $L,W=(13,12)$, respectively, 
where $F_{\textnormal{md}}$ equals $0.11 \pm 0.04$ and $0.20 \pm 0.04$. It should be remembered that the 35 simulated layers had $L_{0}=25$\,m. During CAPT~1 
it was also assumed that each of the fitted layers have $L_{0}=25$\,m. 
As mentioned previously an ROI of $W=1$ has reduced $L_{0}(h)$ information. We can therefore 
not conclude whether an ELT-scale system would show a similar trend to Fig.~\ref{fig:esoWL}. If an ELT-scale 
system is required to measure $L_{0}(h)$ it might be benefitial to have $W>1$.

\subsection{On-sky open-loop AO telemetry optical turbulence profiling}
\label{sec:onSkyResults}

In this section CAPT is demonstrated on-sky using 
open-loop centroid measurements from CANARY. The MOAO capabilities of CANARY have been 
published previously (\citealp{Gendron2014}; \citealp{Vidal2014}; \citealp{Martin2017}). The datasets considered
had four SHWFSs operational, each of which recorded - or were constrained to - 
10,000 centroid measurements from independent NGSs. LGSs are not considered during this section 
because there was not enough open-loop LGS datasets. The frame rate across all observations 
was approximately 150\,Hz. Reference optical turbulence 
profiles were taken using the Stereo-SCIDAR instrument 
that was running at the same time as CANARY but on the INT. The time between WHT 
centroid-data retrieval and a Stereo-SCIDAR optical turbulence profile measurement was 
limited to twenty minutes. It was assumed that the optical turbulence profile would not
drastically alter within this timescale. The constraints on CANARY and the 
availability of Stereo-SCIDAR measurements reduced the number of useful datasets and in 
total, 27 were analysed. These observations were made in July and October of 2014
across 4 and 3 nights, respectively.

\begin{table}
    \centering
    \caption{Covariance matrix and map ROI optical turbulence profiling results from fitting to on-sky 
    CANARY cross-covariance arrays.}
    \label{tab:onSkyResults}
    \begin{tabular}{|l|c|c|}
    \hline
                        & \multicolumn{2}{c|}{$F_{\text{md}}$} \\ \hline
                        & 1 CAPT Iteration   & 15 CAPT Iterations \\ \hline
    Matrix  & 0.46 $\pm$ 0.04 & 0.46 $\pm$ 0.04  \\ \hline
    Map ROI & 0.38 $\pm$ 0.04 & 0.38 $\pm$ 0.04  \\ \hline
    \end{tabular}
\end{table}

\begin{table*}
    \centering
    \caption{Mean absolute values for SHWFS shifts and rotations in the CANARY system. The mean and standard error are calculated from all 27 datasets. 
    SHWFS.4 is the zeroth point for all SHWFS shift misalignments.}
    \label{tab:shwfs_misalignments}
    \begin{tabular}{|l|c|c|c|c|c|c|}
    \hline
            & \multicolumn{3}{c|}{Matrix}                 & \multicolumn{3}{c|}{Map ROI}                \\ \hline
            & Shift Offset in $x$ ($D$) & Shift Offset in $y$ ($D$) & Rotational Offset ($^\circ$) & Shift Offset in $x$ ($D$) & Shift Offset in $y$ ($D$) & Rotational Offset ($^\circ$) \\ \hline
    SHWFS.1 &       0.01           &    0.02            &     1.33                &      0.02              &    0.02                 &         1.80           \\ \hline
    SHWFS.2 &       0.02            &    0.00              &     1.14                &      0.03              &    0.01                  &         2.11            \\ \hline
    SHWFS.3 &       0.01            &    0.00             &     1.08                &      0.01              &    0.01                  &         1.27            \\ \hline
    SHWFS.4 &       0.00            &    0.00               &     1.06                &      0.00                        &    0.00                            &         1.56            \\ \hline
    \end{tabular}
\end{table*}

Covariance matrix and map ROI CAPT optical turbulence profiling was performed on the CANARY 
dataset. Seven evenly-spaced layers were fitted from 0 to 24\,km. The 6 SHWFS 
combinations were fitted to simultaneously. For each observation no layer above 
the $h_{\textnormal{max}}$ altitude bin was fitted. Data from Stereo-SCIDAR was assumed to be a high-resolution measure of the reference optical 
turbulence profile and therefore - as with the simulated profile 
in Section~\ref{sec:simResults} - its measurements were binned to the fitted altitudes to give $C_{n}^{2}(h^{\text{m}})^{\text{r}}$. 
The noise-floor for $C_{n}^{2}(h^{\text{m}})^{\text{r}}dh$ was set at 
$10^{-16}$\,$\textnormal{m}^{1/3}$. 
To be included in the analysis of $F_{\textnormal{md}}$ the fitted layer was required to be centred at or below $h_{\textnormal{max}}$. 
It had to also not extend past the maximum altitude bin of Stereo-SCIDAR.
All remaining measurements were used to calculate the mean value of $F_{\textnormal{md}}$ 
along with its standard error. The results in Table~\ref{tab:onSkyResults} quantify the relation 
between Stereo-SCIDAR and CANARY AO telemetry optical turbulence profiling. These results show $F_{\textnormal{md}}$ after 1 and 
15 CAPT iterations. It is clear from Table~\ref{tab:onSkyResults} that fitting SHWFS misalignments has not noticeably 
improved the system. The mean absolute measure of each SHWFS misalignment is given in Table~\ref{tab:shwfs_misalignments}. 
The standard error on each SHWFS misalignment measurement was negligible. It can be concluded from 
Table~\ref{tab:shwfs_misalignments} that there is no change in $F_{\textnormal{md}}$ 
as the SHWFSs are well-aligned. However, it is important to note that similar misregistrations would significantly impact 
an ELT-scale AO system (see Fig.~\ref{fig:rotOffset}, Fig.~\ref{fig:shiftOffset} and Fig.~\ref{fig:fmd_comp}).  
Covariance matrix fitting found $xx$ and $yy$ vibration artefacts to 
account for an average of $0.011\pm0.001$\,arcsec$^{2}$ and $0.011\pm0.001$\,arcsec$^{2}$ per observation. The covariance map 
ROI measured these vibration artefacts to be $0.012\pm0.001$\,arcsec$^{2}$ and $0.011\pm0.001$\,arcsec$^{2}$.

\begin{figure}
	\includegraphics{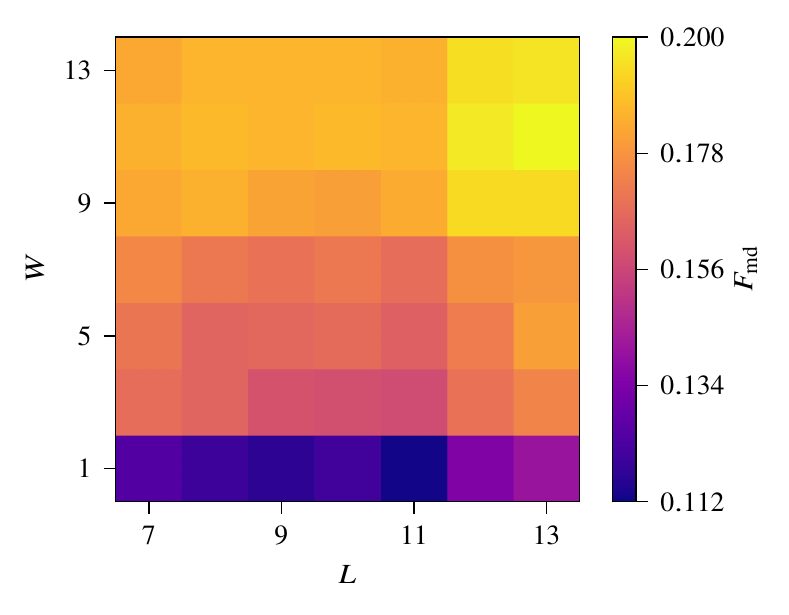}
	\caption{Mean deviation, $F_{\textnormal{md}}$, shown as a function of covariance map ROI length, $L$, and width, $W$.
	The results are from CAPT analysis of simulated NGS CANARY data. The optical turbulence profile reference was the binned form of the 
	35-layer ESO profile.}
	\label{fig:esoWL}
\end{figure}
\begin{figure}
    \includegraphics{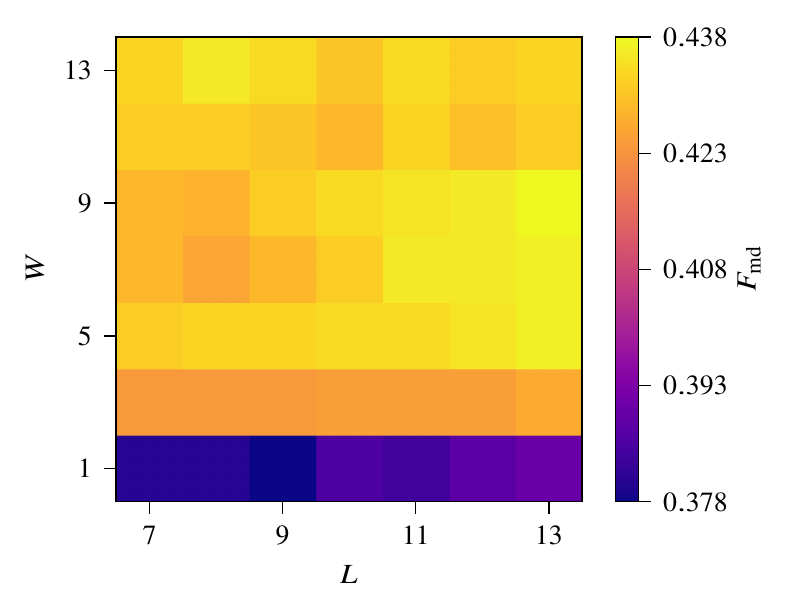}
    \caption{Mean deviation, $F_{\textnormal{md}}$, shown as a function of covariance map ROI length, $L$, and width, $W$.
    The results are from CAPT analysis on-sky NGS CANARY data. The optical turbulence profile reference was the binned form of the 
    corresponding Stereo-SCIDAR observation.}
    \label{fig:scidarWL}
\end{figure}

If the Stereo-SCIDAR measurement is assumed 
to be the reference profile the system is optimised when CAPT is twinned with covariance map ROI data analysis. 
However, the associated uncertainties prevent this from being a definitive conclusion. Although the on-sky results are not as 
accurate as those from Section~\ref{sec:simResults} there are a few additional factors to consider: 
the INT and WHT will both have unique dome and local environmental seeing conditions; 
the telescopes are not observing the same direction; on-sky NGS asterisms will have a range of $\gamma$, $\theta$ and apparent magnitude 
values; the data retrieved for profiling has not been 
recorded at the exact same instant; optical misregistrations such as SHWFS magnification etc. are 
not accounted for.

The investigation into the optimal size of the covariance map ROI was repeated with the on-sky CANARY data. The results 
are shown in Fig.~\ref{fig:scidarWL}. There is a clear resemblance between Fig.~\ref{fig:esoWL} and 
Fig.~\ref{fig:scidarWL}. If binned Stereo-SCIDAR is considered to be the reference optical turbulence profile then both 
simulation and on-sky results agree that the system does not benefit from including baselines at $W>1$. 
To reiterate the discussion in Section~\ref{sec:simResults}: future investigations should proceed with caution as $L_{0}(h)$ 
fitting might require $W>1$. The slight discrepancies between Fig.~\ref{fig:esoWL} and Fig.~\ref{fig:scidarWL} can be attributed to the reasons 
previously listed. The lowest and highest $F_{\textnormal{md}}$ values within 
Fig.~\ref{fig:scidarWL} are $0.38 \pm 0.04$ and $0.44 \pm 0.04$, respectively, and can be found at $L,W=(9,1)$ and $L,W=(13,9)$.

The log-log plot between binned Stereo-SCIDAR and CANARY covariance 
matrix optical turbulence profiling is shown in Fig.\ref{fig:scatter_matrix}. Turbulent layers that are at the noise-floor are 
not shown in the plot but are included in the calculation of $F_{\textnormal{md}}$. The corresponding 
plot for the covariance map ROI is shown in Fig.\ref{fig:scatter_roi}. Both 
Fig.\ref{fig:scatter_matrix} and Fig.\ref{fig:scatter_roi} appear to show a slight bias 
to lower $C_{n}^{2}(h^{\text{m}}>0)^{\text{m}}$ values. The Stereo-SCIDAR observations have a median integrated $r_{0}$ of $0.15$\,m. 
Covariance matrix and map ROI measure median integrated $r_{0}$ to be 0.18 and $0.17$\,m, respectively. At $0$\,km there is an expected bias 
as Stereo-SCIDAR and CANARY are operating out of different telescope domes that are separated by roughly 400\,m.

\section{Computational requirements and efficiency improvements}

\subsection{Computational requirements for analytically generating a covariance matrix and map ROI}
\label{sec:arrayScaling}

We have developed sophisticated algorithms to optimise the efficiency of CAPT. 
In this section we compare the efficiency of CAPT covariance matrix and map ROI data processing.
If SHWFSs with the same dimensions (in units of sub-apertures) are used, 
there is a considerable amount of repetition in an analytically generated covariance matrix e.g. in Fig.~\ref{fig:covMat} 
$x_{1}x_{2}$ is a reflection of $x_{2}x_{1}$ and $x_{1}x_{1}=x_{2}x_{2}$. 
Orthogonal centroid covariance is also repeated for each SHWFS pairing e.g. $x_{1}y_{2}=x_{2}y_{1}$. 
The number of calculations required to analytically generate a covariance matrix that can account for SHWFS misalignments is
\begin{equation}
    \label{eq:nm}
    \begin{aligned}
        N_{\textnormal{m}} =& 3N_{\text{L}}\big(kn_{\textnormal{s}}^{2} + m_{\textnormal{n}}\big),
    \end{aligned}
\end{equation}
where $k$ represents the number of GS combinations. The number of 
sub-apertures and sub-aperture baselines in each SHWFS is given by 
$n_{\textnormal{s}}$ and $m_{\textnormal{n}}$, respectively. Auto-covariance regions are insensitive to 
SHWFS misalignments and so they require as many calculations as they have baselines i.e. auto-covariance regions 
are Toeplitz matrices.

\begin{figure}
    \includegraphics{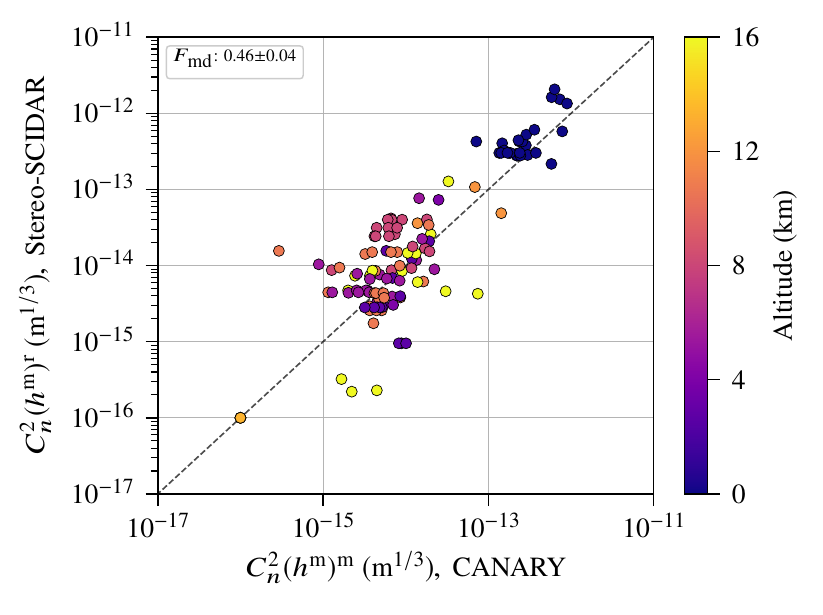}
    \caption{Log-log plot of binned Stereo-SCIDAR and CANARY covariance matrix optical turbulence profile measurements from CAPT. The black dashed line 
    plots where $C_{n}^{2}(h^{\text{m}})^{\text{m}} = C_{n}^{2}(h^{\text{m}})^{\text{r}}$.}
    \label{fig:scatter_matrix}
\end{figure}
\begin{figure}
    \includegraphics{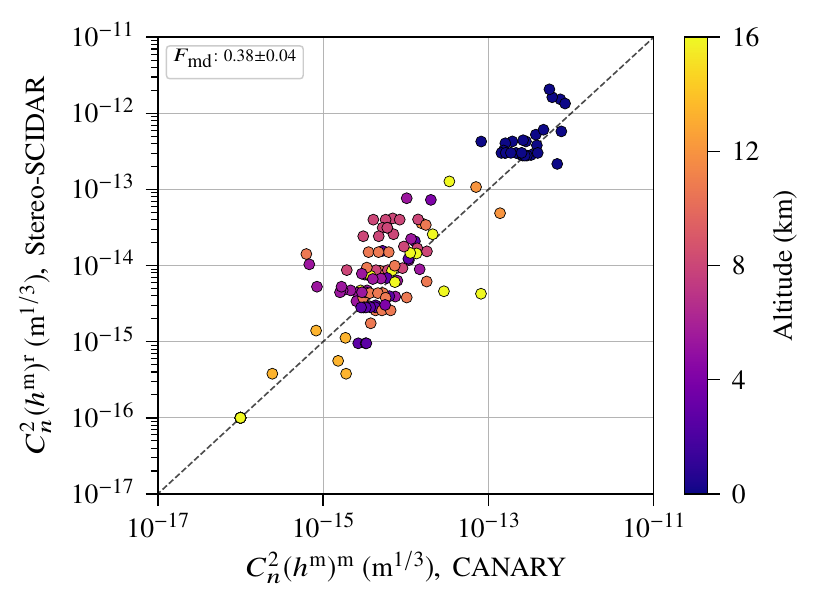}
    \caption{Log-log plot of binned Stereo-SCIDAR and CANARY covariance map ROI optical turbulence profile measurements from CAPT. The black dashed line 
    plots where $C_{n}^{2}(h^{\text{m}})^{\text{m}} = C_{n}^{2}(h^{\text{m}})^{\text{r}}$.}
    \label{fig:scatter_roi}
\end{figure}

\begin{figure}
	\includegraphics{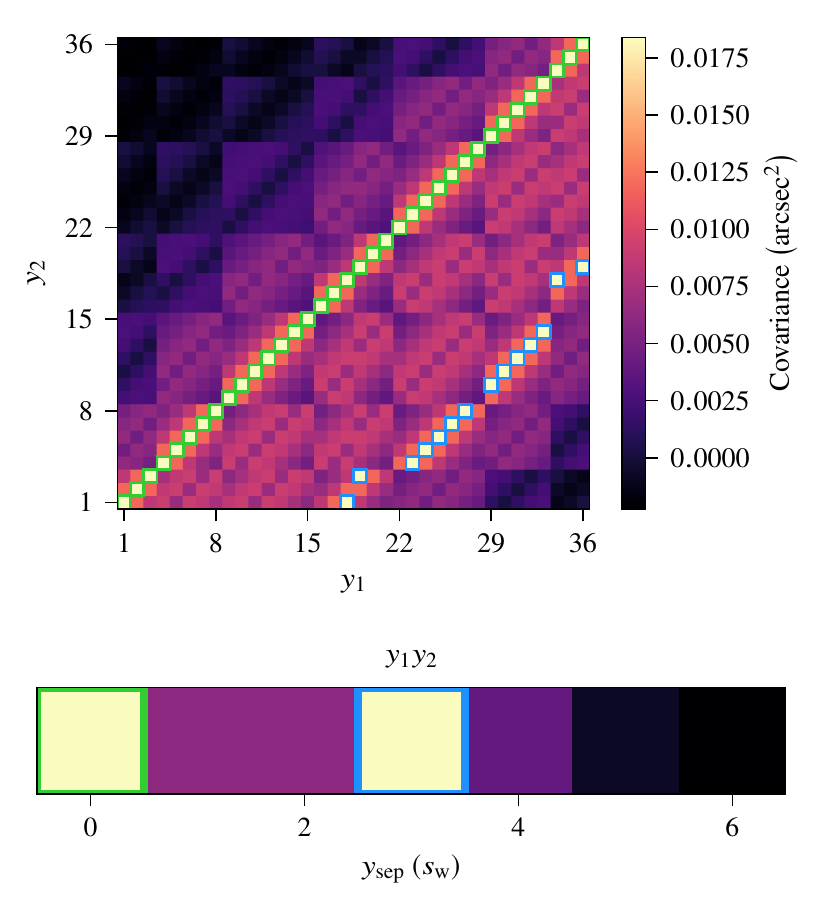}
    \caption{Required data points for generating a $y_{1}y_{2}$ covariance map ROI that can account for SHWFS misalignments. 
    Outlined in green are the 36 analytically generated data points (top figure) that are averaged to baseline 
    $x_{\text{sep}},y_{\text{sep}}=(0,0)$ in the $y_{1}y_{2}$ covariance map ROI from Fig.~\ref{fig:covMapROI} (lower figure). 
    Outlined in blue are the 14 data points that are averaged to the baseline $x_{\text{sep}},y_{\text{sep}}=(0,3)$ in the $y_{1}y_{2}$ covariance map ROI.}
    \label{fig:lit}
\end{figure}

Only specific sub-aperture combinations have to be considered when analytically generating the baseline values within a covariance map ROI (see Fig.~\ref{fig:lit}). 
For a system that can compensate for SHWFS misalignments, each data point within the covariance map ROI must be the baseline average of all 
analytically generated covariance values. This is because misaligned SHWFSs will have baselines comprised of independent sub-aperture 
separations. This averaging is the same process that was outlined in Section~\ref{sec:covMap} however, the model averages 
analytically generated covariance. This is illustrated in Fig.~\ref{fig:lit}. An ROI can be generated for $W>1$ and $L>n_{\text{d}}$ by averaging 
analytical covariance at the additional baselines. It follows that 
only specific covariance data points need to be analytically generated to perform the operation of {\sffamily$\textbf{\text{T}}$} during CAPT~1 (see Section~\ref{sec:l3s} and Section~\ref{sec:capt}). 
Firstly, the covariance map ROI must be analytically generated along 
each value of $\gamma$ for every GS combination. The auto-covariance map ROI 
for each GS combination must also be generated. A further requirement is that for each of these combinations, the covariance of all 
sub-aperture separations that lie along $\gamma$ and $\gamma+\pi$ must be generated i.e. $L=(2n_{\textnormal{d}}-1)$. 
There is no mathematical expression for analytical covariance map ROI common-motion subtraction. 
To perform this operation we have simply created an algorithm that analytically generates only the required covariance data points. 
These data points are then used 
to perform the equivalent operation of equation~\ref{eq:l3s} on the covariance map ROI array.
The number of calculations required to analytically generate a covariance map ROI that can account for SHWFS misalignments 
during CAPT~1 is
\begin{equation}
    \label{eq:nr1}
    \begin{aligned}
        N_{\textnormal{r}} =& 2N_{\text{L}}k^{2}\bigg(2r_{\text{n}} - n_{\text{s}} + \frac{2n_{\textnormal{d}}-1}{k}\bigg).
    \end{aligned}
\end{equation}
In equation~\ref{eq:nr1} $r_{\textnormal{n}}$ is the 
number of sub-aperture separation pairings that are within one axis of the covariance map ROI along $\gamma$ e.g. 
Fig.~\ref{fig:lit} can be used to determine that Fig.~\ref{fig:covMapROI} has $r_{\text{n}}=36+28+21+14+11+6+3=119$. $r_{\text{n}}$ 
is independent of $L$. 
The number of calculations required to generate a 
covariance map ROI in CAPT~2 is $2ks_{\text{n}}$. $s_{\textnormal{n}}$ is the 
number of sub-aperture separation pairings that are within one axis of the covariance map ROI. 
For Fig.~\ref{fig:covMapROI} and Fig.~\ref{fig:lit} $s_{\text{n}}=r_{\text{n}}$. The difference 
between $s_{\text{n}}$ and $r_{\text{n}}$ is that $s_{\text{n}}$ is dependent on $L$.
The number of calculations required to generate a covariance map ROI in CAPT~3 is $2N_{\text{L}}ks_{\text{n}}$. 
If CAPT~3 considers a larger ROI this will increase the value of $s_{\textnormal{n}}$.

Fig.~\ref{fig:calcsCAPT1} shows the number of calculations required to generate a 
single-layer covariance matrix and map ROI during CAPT~1 when misalignments are accounted for. 
These are plotted for CANARY, AOF and ELT-scale AO systems, 
if each were operating a 2, 4 and 6 GS configuration. The ROI for each AO system has $L=n_{\text{d}}$ and $W=1$. 
As the algorithms consider an increased number of baselines, the number of required 
calculations increases at a greater rate for the covariance matrix. 
This makes the reduced computational strain offered by the covariance map ROI especially appealing for ELT-scale instruments. 
During CAPT~1 it should also be noted that as the system 
assumes a known outer scale profile, the covariance calculations 
only need to be performed once. Thereafter CAPT can iteratively fit the overall strength of the optical turbulence profile. 
The plot shown in Fig.~\ref{fig:calcsCAPT1} 
is repeated for CAPT~2 and CAPT~3 in Fig.~\ref{fig:calcsCAPT2} and Fig.~\ref{fig:calcsCAPT3}, respectively. The ROI for 
the AO systems in Fig.~\ref{fig:calcsCAPT2} have $L=n_{\text{d}}$ and $W=1$. In Fig.~\ref{fig:calcsCAPT3} $L=(n_{\text{d}}+1)$ and $W=3$.

In an optically aligned system two-dimensional sub-aperture separation is equal to its baseline value}.
This results in many identical values throughout 
each region of an analytically generated covariance matrix i.e. each pairing of 
SHWFS axes (for example, $x_{1}x_{2}$) is its own Toeplitz matrix. 
This implies that covariance maps can be analytically generated directly and that - as they contain every possible sub-aperture separation 
- they contain every unique covariance value. For example, Fig.~\ref{fig:covMat} was analytically generated for an optically aligned system and therefore, 
all $x_{1}x_{2}$ and $y_{1}y_{2}$ values within Fig.~\ref{fig:covMat} can be found in Fig.~\ref{fig:covMap}.
The number of calculations required in generating a covariance matrix for optically aligned SHWFSs is equal to 
equation~\ref{eq:nm} but with $n_{\textnormal{s}}=\sqrt{m_{\textnormal{n}}}$.
A further implication of this is that for each layer, the generation of a covariance map ROI requires as 
many calculations as it has baselines. For a system that is assumed optically aligned, equation~\ref{eq:nr1} will 
have $r_{\textnormal{n}} = (2n_{\text{d}}-1 + n_{\text{s}})/2$. The number of calculations required to generate a covariance map 
in CAPT~2 becomes $2kLW$.

\begin{figure}
	\includegraphics{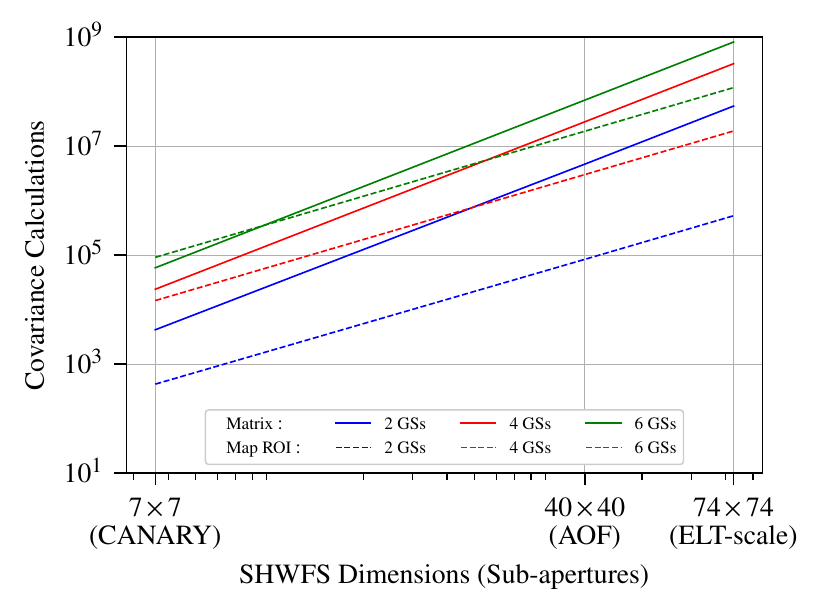}
    \caption{The number of calculations to generate a $N_{\text{L}}=1$
    covariance matrix and map ROI during CAPT~1. The ROI has $L=n_{\text{d}}$ and $W=1$.}
    \label{fig:calcsCAPT1}
\end{figure}

\begin{figure}
	\includegraphics{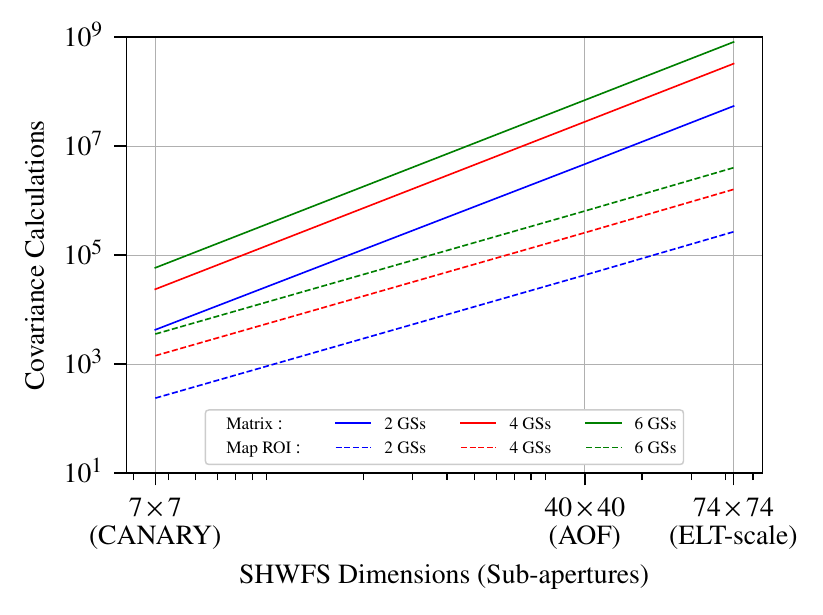}
    \caption{The number of calculations to generate a $N_{\text{L}}=1$
    covariance matrix and map ROI during CAPT~2. The ROI has $L=n_{\text{d}}$ and $W=1$.}
    \label{fig:calcsCAPT2}
\end{figure}

\begin{figure}
	\includegraphics{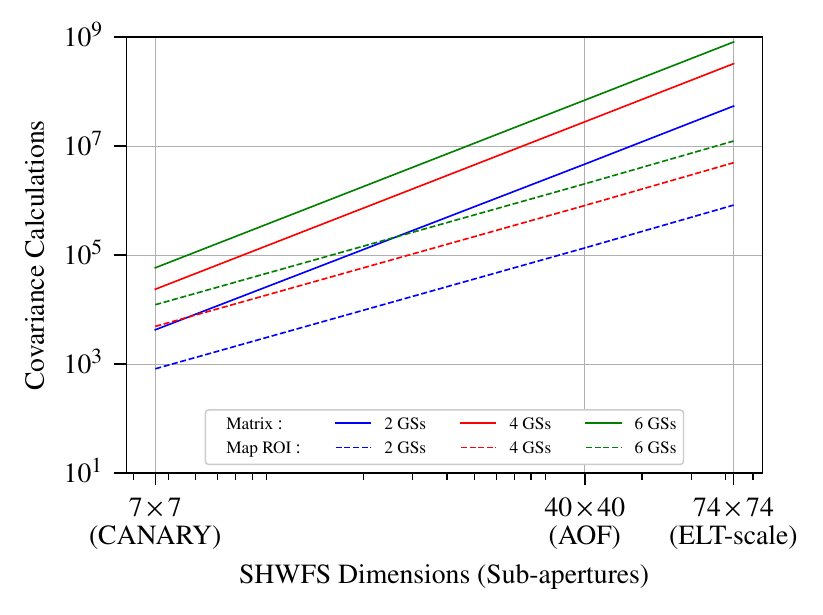}
    \caption{The number of calculations to generate a $N_{\text{L}}=1$
    covariance matrix and map ROI during CAPT~3. The ROI has $L=(n_{\text{d}}+1)$ and $W=3$.}
    \label{fig:calcsCAPT3}
\end{figure}

\subsection{Demonstration of Computational Efficiency}
\label{sec:timing}

The overall computational efficiency of 2-NGS covariance matrix and map ROI optical turbulence profiling is shown here for a CANARY, 
AOF and ELT-scale AO system. The target covariance matrix for each AO system was the $M_{\textnormal{T}}$ from 
Section~\ref{sec:misalignments} with zero SHWFS misalignments. The covariance map ROI was calculated from each $M_{\textnormal{T}}$. 
Seven evenly-spaced turbulent layers were fitted from 0 to 24\,km. As 
each $M_{\text{T}}$ was generated for perfectly aligned SHWFSs both 
covariance matrix and map ROI fitting would recover the exact optical turbulence profile. This meant that the efficiency of CAPT would be 
primarily dependent on computational strain. 
CAPT covariance matrix and map ROI optical turbulence profiling was timed for the CANARY, AOF and ELT-scale 
systems. To start, CAPT assumed perfect optical alignment i.e. the number of analytical covariance calculations was reduced to a minimum 
(see Section~\ref{sec:arrayScaling}). Only CAPT~1 and CAPT~2 can be performed in a system that assumes SHWFS optical alignment. 
To test computational 
requirements against computational efficiency, CAPT did not take advantage of parallel computing. 
The fitting process was operated on a Dell Precision 
Tower 3620 workstation with an Intel Core i7-6700 CPU processor and 64GB of RAM. 
The fitting routines were each repeated 5 times. The standard error of each routine was negligible. 
Table~\ref{tab:CAPT_timing_aligned} summarises the timing results.

\begin{figure}
    \includegraphics{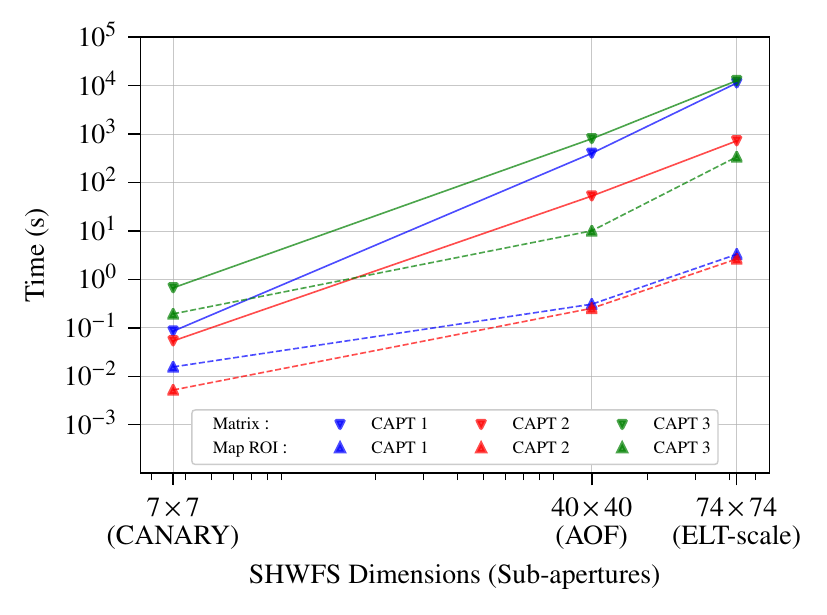}
    \caption{Covariance matrix and map ROI CAPT fitting time when accounting for SHWFS misalignments. The configuration concerns a 2-NGS system and the fitting 
    of 7 layers. Line plots have been overlaid to indicate the general trend across different AO systems.}
    \label{fig:timing}
\end{figure}

The study was repeated but with CAPT accounting for SHWFS misalignments i.e. a system 
that can no longer take advantage of Toeplitz matrix symmetry outside of auto-covariance regions 
(see Section~\ref{sec:arrayScaling}). To test the efficiency against a CAPT 
system that assumes perfect optical alignment, the analytically generated covariance arrays had zero SHWFS misalignments 
during CAPT~1 and CAPT~2. The processing time of each step of CAPT is shown in 
Fig.~\ref{fig:timing}. Table~\ref{tab:CAPT_timing_misaligned} summarises the timing results. The comparison between Table~\ref{tab:CAPT_timing_aligned} 
and Table~\ref{tab:CAPT_timing_misaligned} outlines the loss in computational efficiency when SHWFS misalignments are accounted for.
In Fig.~\ref{fig:timing} all AO systems record CAPT~2 being the fastest stage of CAPT. This is because it is only fitting one layer. 
CAPT~3 is the most computationally demanding stage as each iteration requires the recalculation of sub-aperture 
covariance for 7 layers. In the fitted model the starting point for SHWFS misalignments is zero. $M_{\text{T}}$ was generated 
for perfectly aligned SHWFSs and so CAPT~3 will measure SHWFS misalignments to be zero. Having the starting point for fitting equal 
to the measurement 
does not significantly reduce computing time as the LMA in CAPT~3 still has to perform local optimisation. 
For a CANARY-scale system the efficiency of covariance matrix optical turbulence 
profiling is not an issue. For the AOF the CAPT procedure 
takes over 20\,minutes. For the ELT-scale system covariance matrix CAPT takes almost 7\,hours. The covariance map ROI 
reduces ELT-scale CAPT processing time to under 6 minutes i.e. it improves the computational efficiency of the ELT-scale system 
by a factor of 72. If SHWFS misalignments have already been logged (see Section~\ref{sec:meas_misalign}) then 
the covariance map ROI can perform ELT-scale CAPT optical turbulence profiling (CAPT~1 and CAPT~2) in under 6 seconds. 
As mentioned throughout ELT-scale covariance map ROI data analysis might require $W>1$ for $L_{0}(h)$ fitting in CAPT~1 and CAPT~2. 
This will increase the processing times presented however, using the covariance map ROI for SLODAR data analysis 
will still be the most efficient technique. 
\begin{table}
    \centering
    \caption{CAPT fitting time for covariance matrix and map ROI algorithms that asssume an optically aligned system. 
    The results are for a 2-NGS system where $N_{\text{L}}=7$. Optical turbulence profiling occurs during CAPT~1 and CAPT~2 (CAPT~1+2).}
    \label{tab:CAPT_timing_aligned}
    \begin{tabular}{|l|c|c|} 
    \hline
                & \multicolumn{2}{c|}{Time Taken (s)}                                                            \\ 
    \hline
                & \multicolumn{1}{c|}{Matrix}                      & \multicolumn{1}{c|}{Map ROI}          \\ 
    \hline
    AO System & CAPT~1+2                                         & CAPT~1+2                              \\ 
    \hline
    CANARY    &   0.09                                                    &         0.02                                      \\ 
    \hline
    AOF       &    381.14                                               &       0.04                                      \\ 
    \hline
    ELT-scale   &    $1.12$ $\times$ 10$^{4}$                       &         0.21                                  \\
    \hline
    \end{tabular}
    \end{table}

\section{Conclusions}

To achieve optimal performance forthcoming ELT AO systems will require 
efficient, high-precision measurements of the optical turbulence profile.
This investigation looked at the differences between using the CAPT fitting procedure for covariance matrix and map ROI 
optical turbulence profiling.
Both techniques were tested under SHWFS misalignments at the scale of CANARY, 
AOF and ELT AO systems. SHWFS misalignments were shown to significantly degrade the accuracy of 
optical turbulence profiling. An $N_{\text{L}}=7$ ELT-scale system measured $F_{\textnormal{md}}$ values of approximately 0.6 when SHWFSs 
were misaligned by a rotation of $5^{\circ}$ or a lateral shift of $0.05D$. However, it was also shown that by iterating through CAPT a number of times 
both the covariance matrix and map ROI can measure SHWFS misalignments. We recommend 
15 iterations. Using simulated data for a 4-NGS CANARY system, covariance matrix and map ROI CAPT measured $F_{\textnormal{md}}$ 
to be $0.19\pm0.03$ and $0.13\pm0.03$, respectively. Both techniques were also shown to be applicable to LGS analysis. 
On-sky NGS data from CANARY was used to demonstrate real-world AO telemetry optical turbulence profiling. 
Results were quantified using contemporaneous Stereo-SCIDAR measurements. This was the 
first demonstration of AO telemetry optical turbulence profiling against a dedicated high-resolution optical turbulence profiler. 
For the on-sky optical turbulence profiles, covariance matrix and map ROI CAPT measured $F_{\textnormal{md}}$ 
to be $0.46\pm0.04$ and $0.38\pm0.04$, respectively. Measured SHWFS misalignments from on-sky data indicated that the CANARY 
system was well-aligned. Results from simulated and on-sky CANARY data showed that the optimal 
covariance map ROI has $W=1$. It was also shown that compared to the covariance matrix, 
the covariance map ROI improves the efficiency of an ELT-scale system 
by a factor of $72$.

\begin{table}
    \centering
    \caption{CAPT fitting time for covariance matrix and map ROI algorithms that are accounting for SHWFS 
    misalignments. The results are for a 2-NGS system where $N_{\text{L}}=7$.  Optical turbulence profiling occurs during CAPT~1 
    and CAPT~2 (CAPT~1+2). SHWFS misalignments are fitted during CAPT~3.}
    \label{tab:CAPT_timing_misaligned}
    \begin{tabular}{|l|c|c|c|c|} 
    \hline
              & \multicolumn{4}{c|}{Time Taken (s)}                                                            \\ 
    \hline
              & \multicolumn{2}{c|}{Matrix}                      & \multicolumn{2}{c|}{Map ROI}          \\ 
    \hline
    AO System & CAPT~1+2                                         & CAPT~3 & CAPT~1+2                             & CAPT~3  \\ 
    \hline
    CANARY    &   0.1                                      &  0.67                          &       0.02                                    &    0.19      \\ 
    \hline
    AOF       &    454.17                                  &  801.41                        &       0.56                                    &    10.06     \\ 
    \hline
    ELT-scale   &    $1.21$ $\times$ 10$^{4}$                &   $1.27$ $\times$ 10$^{4}$     &       5.93                                    &    338.52    \\
    \hline
    \end{tabular}
\end{table}

The covariance map ROI is capable of measuring SHWFS misalignments using an iterative CAPT approach. 
It also outperforms the accuracy of covariance matrix optical turbulence profiling. In addition, 
the covariance map ROI improves the efficiency of an ELT-scale AO system by almost 
two orders of magnitude. We conclude from this study that AO telemetry optical 
turbulence profiling should twin CAPT with covariance map ROI data analysis.

\section*{Acknowledgements}

All of the plots were rendered in Matplotlib (\citealp{Hunter2007}).
The William Herschel Telescope and the Isaac Newton Telescope are operated on the island of La Palma by the 
Isaac Newton Group in the Spanish Observatorio del Roque de los Muchachos of the Instituto de Astrof\'isica de Canarias. 
The work carried out at both of these telecopes was supported by the Science and Technology 
Funding Council, UK (grant CG ST/P000541/1) and the European Commission 
(FP7 Infrastructures OPTICON grant 226604, 312430, and H2020 grant 730890). 
The raw data used in this publication is available
from the authors. We thank the UK Programme for the 
European Extremely Large Telescope (ST/N002660/1) for supporting this research. We would 
also like to thank the reviewer of this publication for their constructive comments.





\begin{thebibliography}{99}


\bibitem[\protect\citeauthoryear{Butterley, Wilson \& Sarazin}{2006}]{Butterley2006}
Butterley T., Wilson R. W., Sarazin M., 2006, MNRAS, 369, 835-845

\bibitem[\protect\citeauthoryear{Cort\'es et al.}{2012}]{Cortes2012}
Cort\'es A., Neichel B., Guesalaga A., Osborn J., Rigaut F., Duzman D., 2012, MNRAS, 427, 2089-2099


\bibitem[\protect\citeauthoryear{Gendron et al.}{2014}]{Gendron2014}
Gendron E., Vidal F., Brangier M., Morris T., Hubert Z., Basden A., Rousset G., Myers R., Chemla F., Longmore A., 
Butterley T., Dipper N., Dunlop C., Geng D., Gratadour D., Henry D., Laporte P., Looker N., Perret D., Sevin A., 
Talbot G., Younger E., 2011, Astronomy \& Astrophysics, 529

\bibitem[\protect\citeauthoryear{Guesalaga et al.}{2014}]{Guesalaga2014}
Guesalaga A., Neichel B., Cortes A., B\'echet C., Guzm\'an D., 2014, MNRAS, 440, 1925-1933

\bibitem[\protect\citeauthoryear{Guesalaga et al.}{2017}]{Guesalaga2017}
Guesalaga A., Neichel B., Correia C. M., Butterley T., Osborn J., Masciadri E., Fusco T., Sauvage J.-F., 2017, MNRAS, 465, 1984-1994

\bibitem[\protect\citeauthoryear{Hardy}{1998}]{Hardy1998}
Hardy J., 1998, Oxford University Press, Adaptive Optics for Astronomical Telescopes

\bibitem[\protect\citeauthoryear{Hunter}{2007}]{Hunter2007}
Hunter J. D., Matplotlib: A 2D graphics environment, Computing In Science \& Engineering, Vol. 9, 90-95, 2007

\bibitem[\protect\citeauthoryear{Jones et al.}{}]{Jones}
Jones E., Oliphant T., Peterson P, and Others, SciPy: Open source scientific tools for Python

\bibitem[\protect\citeauthoryear{Martin et al.}{2017}]{Martin2017}
Martin O. A., Gendron \'{E}., Rousset G., Gratadour D., Vidal F., Morris T. J., Basden A. G., Myers R. M., Correia C. M., Henry D., 2017, Astronomy \& Astrophysics, 598

\bibitem[\protect\citeauthoryear{Martin et al.}{2016}]{Martin2016}
Martin O., Correia C. M., Gendron E., Rousset G., Vidal F., Morris T. J., Basden A. G., Myers R. M., Ono Y. H., Neichel B., Fusco T., 2016, SPIE, Adaptive Optics Systems V, 9909

\bibitem[\protect\citeauthoryear{Martin et al.}{2012}]{Martin2012}
Martin O., Gendron E., Rousset G., Vidal F., 2012, SPIE Proceedings, Adaptive Optics Systems III, 8447

\bibitem[\protect\citeauthoryear{Masciadri \& Lascaus}{2012}]{Masciadri2012}
Masciadri E., Lascaus F., 2012, SPIE Proceedings, Adaptive Optics Systems III, 8447

\bibitem[\protect\citeauthoryear{Morris et al.}{2014}]{Morris2014}
Morris T., Gendron E., Basden A., Martin O., Osborn J., Henry D., Hubert Z., Sivo G., Gratadour D., 
Chemla F., Sevin A., Cohen M., Younger E., Vidal F., Wilson R., Butterley T., Bitenc U., Reeves A., Bharmal N., 
Raynaud H., Kulcsar C., Conan J., Huet J., Perret D., Dickson C., Atkinson D., Baillie T., Longmore A., 
Todd S., Talbot G., Morris S., Rousset G., Myers R., 2014, SPIE Proceedings, Adaptive Optics Systems IV, 8447

\bibitem[\protect\citeauthoryear{Neichel et al.}{2009}]{Neichel2009}
Neichel B., Fusco T., Conan J-M., 2009, OSA, 26, A219-A235

\bibitem[\protect\citeauthoryear{Ono et al.}{2016}]{Ono2016}
Ono Y. H., Correia C. M., Andersen D. R., Lardi\`ere O., Oya S., Akiyama M., Jackson K., Bradley C., 2016, MNRAS

\bibitem[\protect\citeauthoryear{Osborn et al.}{2015}]{Osborn2015}
Osborn J., Butterley T., Perera S., Fohring D., Wilson R. W., 2015, AO4ELT4, 1

\bibitem[\protect\citeauthoryear{Peng et al.}{2018}]{Peng2018}
Peng J., Osborn J., Letian K., Laidlaw D., Li C., Farley O., Xue G., 2018, MNRAS, 480, 2466-2474

\bibitem[\protect\citeauthoryear{Petit et al.}{2009}]{Petit2009}
Petit C., Conan J., Kulcsar C., Raynaud H., 2009, Journal of the Optical Society of America, 26, 1307-1325

\bibitem[\protect\citeauthoryear{Reeves}{2016}]{Reeves2016}
Reeves, 2016, SPIE Proceedings, Adaptive Optics Systems V, 9909

\bibitem[\protect\citeauthoryear{ESO}{2015}]{eso2015}
Relevant Atmospheric Parameters for E-ELT AO Analysis and Simulations, ESO-258292, Version 2

\bibitem[\protect\citeauthoryear{Roddier}{1999}]{Roddier1999}
Roddier F., 1999, Cambridge University Press, Adaptive Optics in Astronomy

\bibitem[\protect\citeauthoryear{Shepherd et al.}{2013}]{Shepherd2013}
Shepherd H. W., Osborn J., Wilson R., Butterley T., Avila R., Dhillon V. S., Morris T., 2013, MNRAS, 437, 3568-3577

\bibitem[\protect\citeauthoryear{van der Walt, Colbert \& Ga\"{e}l}{2011}]{Vanderwalt2011}
van der Walt S., Chris Colbert S., Ga\"{e}l, V., The NumPy array: a structure for efficient numerical computation, Computing in Science \& Engineering, Vol. 13, 22-30, 2011

\bibitem[\protect\citeauthoryear{Vernin et al.}{1973}]{Vernin1973}
Vernin J., Roddier F., 1973, Journal of the Optical Society of America, 63, 270-273

\bibitem[\protect\citeauthoryear{Vidal et al.}{2010}]{Vidal2010}
Vidal F., Gendron E., Brangier M., Sevin A., Rousset G., Hubert Z., 2010, AO4ELT1, 07001 

\bibitem[\protect\citeauthoryear{Vidal et al.}{2014}]{Vidal2014}
Vidal F., Gendron {\'E}., Rousset G., Morris T., Basden A., Myers R., Brangier M., Chemla F., Dipper N.,
Gratadour D., Henry D., Hubert Z., Longmore A., Martin O., Talbot G., Younger E., 2014, Astronomy \& Astrophysics, 569

\bibitem[\protect\citeauthoryear{Villecroze et al.}{2012}]{Villecroze2012}
Villecroze R., Fusco T., Bacon R., Madec PY., 2012, SPIE Proceedings, Adaptive Optics Systems III, 8447

\bibitem[\protect\citeauthoryear{Wilson}{2002}]{Wilson2002}
Wilson R., 2002, MNRAS, 337, 103-108

\bibitem[\protect\citeauthoryear{Ziad et al.}{2004}]{Ziad2004}
Ziad A., Sch\"ock M., Chanan G., Troy M., Dekany R. G., Lane B. F., Borgnino J., Martin F., 2004, Appl. Opt., 43, 2316-2324
  


\end{thebibliography}



\bsp	
\label{lastpage}
\end{document}